\documentclass[final]{IEEEtran}

\usepackage{amsmath,amsfonts,amssymb,amsopn}
\usepackage{graphicx}
\usepackage{xcolor}
\usepackage{cite}
\usepackage{xspace}
\usepackage{subfig} 
\usepackage{mathtools} 
\usepackage{url}
\usepackage{algorithm}
\usepackage{algpseudocode}
\usepackage{ifpdf}

\ifpdf
  \usepackage[pdftex,colorlinks=true,breaklinks=true,citecolor=blue]{hyperref}
\else
  \usepackage[ps2pdf,colorlinks=true,citecolor=blue]{hyperref}
\fi

\usepackage{float}
\floatname{algorithm}{Procedure}

\def\NoNumber#1{{\def\alglinenumber##1{}\State #1}\addtocounter{ALG@line}{-1}} 

\ifCLASSOPTIONonecolumn
    
    \renewcommand{\baselinestretch}{2}
    \usepackage[margin=25mm]{geometry}
\else
    
\fi

\newcommand{\bv}[1]{\boldsymbol{#1}}
\newcommand{\dfn}{\triangleq}
\newcommand{\untsph}{\mathbb{S}^{2}} 

\newcommand{\shc}[3]{({#1})_{#2}^{#3}}
\newcommand{\lsph}{L^2(\untsph)}
\newcommand{\conj}[1]{\overline{#1}} 

\newcommand{\lsphL}[1]{\mathcal{H}_{#1}}

\newcommand{\intsph}{\int_{\untsph}}

\newcommand{\intphi}{\int_{0}^{2\pi}}

\newcommand{\figref}[1]{Fig.\,\ref{#1}}

\newcommand{\appref}[1]{Appendix\,\ref{#1}}
\newcommand{\secref}[1]{Section\,\ref{#1}}

\newcommand{\tilf}{\tilde f}
\newcommand{\fs}{\ensuremath{{}_s f}}
\newcommand{\fso}{\ensuremath{{}_0 f}}
\newcommand{\matlab}{\texttt{MATLAB}}

\newcommand{\clang}{\texttt{C}}

\newcommand{\plms}{\tilde{P}}

\newcommand{\ylms}[2]{\ensuremath{{}_s Y_{#1}^{#2}}}
\newcommand{\ylmso}[2]{\ensuremath{{}_0 Y_{#1}^{#2}}}

\newcommand{\nps}{\ensuremath{N_\textrm{O}}}

\DeclarePairedDelimiterX\innerp[2]{\langle}{\rangle}{#1,#2}

\newtheorem{remark}{Remark}

\graphicspath{{figs/},{pdfs/},{pdf/},{Figures/},{Authors/}}

\begin{document}
\title{An Optimal-Dimensionality Sampling Scheme on the Sphere with Fast Spherical Harmonic Transforms}
%
\author{%
Zubair~Khalid,~\IEEEmembership{Member,~IEEE}, Rodney A. Kennedy,~\IEEEmembership{Fellow,~IEEE}, and Jason~D.~McEwen,~\IEEEmembership{Member,~IEEE}
 \thanks{Z.~Khalid and R.~A.~Kennedy are with
   the Research School of Engineering, College of Engineering and
   Computer Science, The Australian National University, Canberra,
   Australia. J.~D.~McEwen is with the Mullard Space Science Laboratory, University College London, Surrey RH5 6NT, UK.}
\thanks{Z.~Khalid and R.~A.~Kennedy are supported by the
  Australian Research Council's Linkage Projects funding scheme (Project no. LP100100588). J.~D.~McEwen is supported in part by a Newton International Fellowship from the Royal Society and the British Academy.}

\thanks{E-mail: zubair.khalid@anu.edu.au, rodney.kennedy@anu.edu.au, jason.mcewen@ucl.ac.uk }
}
\maketitle

\begin{abstract}

We develop a sampling scheme on the sphere that
permits accurate computation of the spherical harmonic transform
and its inverse for signals band-limited at $L$ using only $L^2$ samples.  We obtain the optimal number of samples given by the degrees of freedom of the signal in harmonic space.  The number of samples required in our scheme is a factor of two or four fewer than existing techniques, which require either $2L^2$ or $4L^2$ samples.  We note, however, that we do not recover a sampling theorem on the sphere, where spherical harmonic transforms are theoretically exact.  Nevertheless, we achieve high accuracy even for very large band-limits.  For our optimal-dimensionality sampling scheme, we develop a fast and accurate algorithm to compute the spherical harmonic transform (and inverse), with computational complexity comparable with existing schemes in practice. We conduct numerical experiments to study in detail the stability, accuracy and computational complexity of the proposed transforms. We also highlight the advantages of the proposed sampling scheme and associated transforms in the context of potential applications.
\end{abstract}


\section{Introduction}

\PARstart{S}{ignals} are inherently defined on the sphere in a variety of fields of science and engineering. These include geodesy~\cite{Wieczorek:2007}, cosmology~\cite{Spergel:2007}, computer graphics~\cite{Ng:2004}, medical imaging~\cite{Chung:2007}, astrophysics~\cite{Jarosik:2010}, quantum chemistry~\cite{Ritchie:1999}, wireless communication~\cite{Pollock:2003}, acoustics~\cite{Zhang:2012} and planetary science~\cite{Audet:2011}, to name a few. In signal processing analysis on the sphere~(e.g.,\cite{Driscoll:1994,McEwen:2011,Antoine:1998:wavelets,Kennedy:2011,Sadeghi:2012,Yeo:2008,Khalid:2011,McEwen:2006,mcewen:fcswt,McEwen:2008,Wiaux:2005,Starck:2006,Marinucci:2008,Wiaux:2008,Wieczorek:2005,Simons:2006,Audet:2011,Audet:2014,Leistedt:2012,Kennedy-book:2013}) the signal is often analysed in both the spherical~(spatial) domain and harmonic~(spectral) domain.  The transformation from spatial to spectral is through the spherical harmonic transform~(SHT)~(see, e.g., \cite{Colton:1998,Sakurai:1994,Driscoll:1994,McEwen:2011,Kennedy-book:2013}), which is the well-known counterpart of the Fourier transform. For example, analysis of signals in the spectral domain through the SHT has been instrumental in refining the standard cosmological model and in the study of the anisotropies in the cosmic microwave background~(CMB) \cite{planck2013-p01}. Consequently, the ability to compute the SHT of a signal is of significant importance. Furthermore, since data-sets on the sphere can be of considerable size\cite{planck2013-p01}, and the cost of acquiring samples on the sphere can be large\cite{Zhang:2012,Tuch:2004}, the computation of the SHT of the signal should require the minimum possible number of samples, and be computationally accurate and efficient.

The development of sampling schemes on the sphere and computationally efficient methods to compute the spherical harmonic transform from samples has been
investigated extensively in the literature\cite{Driscoll:1994,Sneeuw:1994,Mohlenkamp:1999,Gorski:2005,Crittenden:1998,Healy:2003,Wiaux:2005:2,Blais:2006,Kostelec:2008,mcewen:fsht,Huffenberger:2010,McEwen:2011}. Sampling schemes and their associated SHT computational methods can be evaluated by three key criteria:  (1) the number of samples, defined as the spatial dimensionality;
(2) the computational complexity; and (3) the numerical accuracy.
In this work, we propose a sampling scheme for band-limited signals on the sphere which requires the same number of samples on the sphere as the number of the degrees of freedom of the signal in harmonic space. Furthermore, we develop an accurate method to compute the SHT 
with complexity scaling, in practice, comparable with the existing schemes. We first review the developments made in the literature followed by a summary of the contributions of this paper.

\subsection{Relation to Prior Work}

Among existing sampling schemes in the literature, iso-latitude sampling schemes~(e.g.,\cite{Driscoll:1994,Crittenden:1998,Healy:2003,Gorski:2005,Wiaux:2005:2,Kostelec:2008,Huffenberger:2010,McEwen:2011,Sneeuw:1994,Blais:2006}), where the samples along longitude are taken over iso-latitude rings~(annuli), enable a separation of variables in the computation of the SHT, which results in a reduction in computational complexity. For the computation of spherical harmonic transforms, sampling theorems have been constructed \cite{Driscoll:1994,Healy:2003,Huffenberger:2010,McEwen:2011}, which lead to theoretically exact SHTs, in addition to other numerical approaches, such as approximate quadrature~\cite{Crittenden:1998,Gorski:2005}), least squares~\cite{Sneeuw:1994,Blais:2006} or spherical designs~\cite{An:2010,Bannai:2009}, which nevertheless often lead to accurate transforms. We focus our attention on an iso-latitude sampling scheme that facilitates accurate computation of the SHT of a signal that is band-limited at $L$ (formally defined in \secref{sec:harmonic_standard_expansion}). For the accurate computation of the SHT of a signal band-limited at $L$, the optimal spatial dimensionality, denoted by $\nps$, attainable by any sampling scheme on the sphere is given by $\nps = L^2$, which is the number of degrees of freedom in harmonic space.  Existing schemes either require $2L^2$ or $4L^2$ samples and therefore do not achieve the optimal spatial dimensionality. We refer the reader to \cite{McEwen:2011} for a more comprehensive review of existing sampling schemes.

An exact method to compute the SHT, based on a sampling theorem on the sphere, was developed by Driscoll and Healy in \cite{Driscoll:1994} exploiting an equiangular sampling comprised of $2L$ iso-latitude rings of points, where the number of points along longitude, in each ring, is the same and equal to $2L-1$. Thus the spatial dimensionality of Driscoll and Healy sampling scheme is \mbox{$\sim 4 L^2$}. The well-known Gauss-Legendre quadrature on the sphere~\cite{Skukowsky:1986,Doroshkevich:2005} may also be used to
construct a sampling theorem and exact SHT from \mbox{$\sim 2 L^2$} samples on the sphere.  The placement of $L$ iso-latitude rings is given by the roots of the Legendre polynomials of order $L$, as dictated by Gauss-Legendre quadrature, and the number of points in each ring remains $2L-1$.  More recently, a new sampling theorem based on an equiangular sampling scheme has been proposed by McEwen and Wiaux, which achieves spatial dimensionality \mbox{$\sim 2 L^2$}~\cite{McEwen:2011}, and requries $3(L-1)$ fewer samples than the Gauss-Legendre approach. For all of these sampling schemes the associated spherical harmonic transforms (that are stable and do not require precomputation) have computational complexity of $O(L^3)$. We note that an algorithm was developed for the Driscoll and Healy sampling theorem \cite{Healy:2003} with complexity $O(L^2 \log^2L)$, but it requires $O(L^3)$ precomputation and storage. The precomputation is practicable for applications in acoustics~\cite{Zhang:2012}, quantum chemistry~\cite{Ritchie:1999}, medical imaging~\cite{Chung:2007}, where the band-limit is of the order $10-10^2$. However, the precomputation becomes infeasible for applications in astrophysics~\cite{Jarosik:2010} and cosmology~\cite{Spergel:2007}, where the band-limit is of the order $10^3-10^4$, as the precomputation requires $1.2$GB of storage for \mbox{$L=1024$}~\cite{Wiaux:2005:2}, which scales to approximately $77$GB for the band-limit $L=4096$~\cite{Huffenberger:2010,McEwen:2011}.

SHTs using least squares approaches have also been developed for equiangular sampling schemes~\cite{Sneeuw:1994,Blais:2006}, which require $2L^2$ samples and achieve good accuracy. However, a naive application of least squares is computationally inefficient since the computational complexity to compute SHTs scales as $O(L^6)$.  However, a separation of variables can be employed to reduce the complexity to $O(L^4)$~\cite{Blais:2006}. Moreover, a least squares method can also be developed to compute SHTs using the optimal number~($\nps = L^2$) of spatial samples~(either regularly or irregularly) distributed over the sphere, but the complexity of such method scales with $O(L^6)$ and therefore least squares becomes computationally infeasible even for small band-limits. We note that $O(L^6)$ complexity of the least squares approach using $\nps$ samples cannot be reduced by a separation of variables since aliasing errors are introduced if the number of samples along longitude in each iso-latitude ring is smaller than $2L-1$. Furthermore, the accuracy and stability of the least squares approach cannot be guaranteed. For example, the reconstruction error in the least squares system using $\nps$ samples is $10^{-6}$~(obtained through numerical experiments on the regular grid) for band-limit $L=64$ and the error grows with the band-limit.

We also note the work on spherical designs for computing integrals over the sphere using quadrature based on uniform weighting~(see \cite{An:2010,Bannai:2009} for a comprehensive review). Is it numerically, but rigorously, proved that the computation of the SHT of a band-limited signal using spherical designs can be performed with $4L^2$ samples~\cite{Chen:2011}. It is also conjectured that in fact $2L^2$ samples may be used~\cite{Sloan:2009}, although this is not proved. Spherical designs with $4L^2$ samples have been constructed successfully for band-limit up to only $L=100$~\cite{Chen:2011}.

\subsection{Contributions}

A summary of the contributions of this paper are as follows.
\begin{itemize}
\item We develop a sampling scheme on the sphere that permits accurate computation of the SHT for band-limited signals, attaining the optimal spatial dimensionality of $\nps = L^2$.
\item We develop a computationally efficient method to compute the SHT and its inverse (called the forward and inverse SHT in the sequel) using our optimal spatial dimensionality sampling scheme, which has complexity with scaling, in practice, comparable to the existing methods, which do not achieve optimal spatial dimensionality.  Furthermore, once the sample positions are determined no additional precomputation is required.
\item We characterize the numerical accuracy and computational complexity of our proposed SHT as a function of the band-limit parameter demonstrating its feasibility on large real-world data-sets.
\end{itemize}

\subsection{Paper Organization}

We review the mathematical background and harmonic analysis on the sphere in \secref{sec:models}. In \secref{sec:nsht}, we propose the sampling scheme on the sphere, which achieves optimal spatial dimensionality, and present the associated novel SHT. We evaluate the numerical accuracy of the proposed SHT, present its computational complexity analysis and outline potential applications in \secref{sec:analysis}. The concluding remarks are made in \secref{sec:conclusions}.

\section{Preliminaries}\label{sec:models}
In this section we review the mathematical background for signals and harmonic analysis on the sphere.

\subsection{Signals on the Sphere}
\label{sec:models:signals:sphere}
In this work, we consider square integrable complex functions
of the form $f(\theta,\phi)$, defined on unit sphere $\mathbb{S}^2 \triangleq
\{\mathbf{u} \in \mathbb{R}^3 \colon |\mathbf{u}| = 1  \}$, where $|\cdot|$ denotes the Euclidian norm, $\theta \in [0, \pi]$ denotes the co-latitude and $\phi\in [0, 2\pi)$ denotes the longitude. The inner product of two
functions $f$ and $h$ defined on $\mathbb{S}^2$ is defined
as~\cite{Kennedy:2011}
\begin{align}\label{eqn:innprd}
\langle f, h \rangle \triangleq  \int_{\mathbb{S}^2}
f(\theta,\phi) \overline {h(\theta,\phi)}
\,\sin\theta\, d\theta\, d\phi,
\end{align}
where $\overline{(\cdot)}$ denotes the complex conjugate,
$\sin\theta\,d\theta\,d\phi$ denotes the differential area element on the sphere and the
integration is carried out over the sphere, that is, $\int_{\mathbb{S}^2} = \int_{\theta=0}^{\pi}\int_{\phi=0}^{2\pi}$ . With the inner
product in \eqref{eqn:innprd}, the space of square integrable complex
valued functions on the sphere forms a complete Hilbert space
$\lsph$. Also, the inner product in \eqref{eqn:innprd} induces a norm
$\|f\| \triangleq\langle f,f \rangle^{1/2}$. We refer to the functions
with finite induced norm as signals on the sphere.

\subsection{Harmonic Analysis on the Sphere}
\label{sec:harmonic_standard_expansion}
The Hilbert space $\lsph$ is separable and the spherical harmonic functions (or spherical harmonics for short) $Y_{\ell}^m(\theta, \phi)$~\cite{Kennedy-book:2013,Sakurai:1994,Colton:1998} of all degrees $\ell \ge 0$ and orders $-\ell\le m \le \ell$ form the archetype complete orthonormal set of basis functions. By completeness, any signal $f\in \lsph$ can be expanded as
\begin{equation}
\label{Eq:f_expansion}
    f(\theta,\phi)=\sum_{{\ell}=0}^{\infty}\sum_{m=-{\ell}}^{\ell} \shc{f}{\ell}{m}  Y_{\ell}^m(\theta,\phi),
\end{equation}
where
\begin{equation}\label{Eq:fcoeff}
    \shc{f}{\ell}{m}\dfn\innerp[\big]{f}{Y_{{\ell}}^m} =
        \intsph f(\theta,\phi)\conj {Y_{\ell}^m(\theta,\phi)} \,\sin\theta\,d\theta\,d\phi
\end{equation}
is the spherical harmonic coefficient of degree $\ell$ and order $m$. Some background properties of spherical harmonics used in this work are given in \appref{App:maths}.

The signal $f\in\lsph$ is defined to be band-limited at degree $L$ if $\shc{f}{\ell}{m} = 0$ for $\ell \ge L$. The set of \emph{bandlimited} signals forms an $L^2$ dimensional subspace of $\lsph$, which is denoted by $\lsphL{L}$.

\subsection{Spin Functions on the Sphere}

The spin $s$ functions on the sphere, denoted by $\fs\in\lsph$ and parameterized by integer spin $s$, are special functions which are defined by their behaviour under local rotations. The local rotation by $\gamma$, rotates the spin function $\fs(\theta,\phi)$ by $\gamma$ in the tangent plane formed at a point on the sphere characterized by $\theta$ and $\phi$~\cite{Zaldarriaga:1997}. Under such rotation, the rotated spin function $\fs'$ is related to the original spin function $\fs$ through
\begin{equation}\label{Eq:spin_property}
\fs'(\theta,\phi) = e^{-is\gamma} \fs(\theta,\phi).
\end{equation}
For spin parameter $s=0$, the spin function $\fs$ becomes the standard~(non-spin or scalar) function $f$~(defined in \secref{sec:models:signals:sphere} with harmonic expansion in \secref{sec:harmonic_standard_expansion}) on the sphere, that is $\fso=f$.

The spin spherical harmonics~(sometimes also referred to as spin weighted spherical harmonics), denoted by $\ylms{\ell}{m}$ and defined for degree $\ell$, order $|m|\leq \ell$ and spin $|s|\leq \ell$ form a complete set of basis functions for spin $s$ functions on the sphere~(see \appref{App:maths} for the definition of $\ylms{\ell}{m}$). Spin spherical harmonics also satisfy the property in \eqref{Eq:spin_property} and therefore serve as a more suitable choice of basis functions for the following expansion of spin functions
\begin{align}\label{Eq:f_spin_expansion}
\fs(\theta,\phi) = \sum_{\ell=0}^{\infty}\sum_{m=-\ell}^{\ell} {\shc{\fs}{\ell}{m}}\, { \ylms{\ell}{m}(\theta,\phi)},
\end{align}
where
\begin{align}
\nonumber
{\shc{\fs}{\ell}{m}} \dfn \innerp{\fs}{\ylms{\ell}{m}}.
\end{align}
The spin function $\fs$ is said to be band-limited at $L$ if ${\shc{\fs}{\ell}{m}}=0$ for all $\ell\geq L$. The set of such band-limited spin functions $\fs$ for each $s$ form a subspace of $\lsph$ and is denoted by ${}_s\lsphL{L}$. Furthermore, we note that $\ylmso{\ell}{m} = Y_\ell^m$, that is, the spin spherical harmonic become the standard spherical harmonic for $s=0$. In the sequel, any reference to a function~(or signal) and spherical harmonic means finite energy scalar function~($\fso=f$) on the sphere and scalar spherical harmonic~($\ylmso{\ell}{m} = Y_\ell^m$) respectively, unless otherwise explicitly stated that the signal or spherical harmonic under consideration is spin weighted.

\section{Optimal Sampling Scheme and Novel Spherical Harmonic Transform}
\label{sec:nsht}

We first consider the spherical harmonic transform~(SHT) of band-limited scalar functions $f \in \lsphL{L}$, for which the summation over degree $\ell$ in \eqref{Eq:f_expansion} is truncated to $L-1$, that is,
\begin{equation}\label{Eq:f_expansion_bandlimited}
f(\theta,\phi)=\sum_{{\ell}=0}^{L-1}\sum_{m=-{\ell}}^{\ell} \shc{f}{\ell}{m}  Y_{\ell}^m(\theta,\phi).
\end{equation}
We first present our sampling scheme for the discretization of a band-limited signal on the sphere. Then we develop the proposed forward SHT which determines the spherical harmonic coefficients $\shc{f}{\ell}{m}$ for $ 0 \leq \ell < L$ and $|m| \leq \ell$ using the discretized signal. We also present the inverse SHT to determine the signal efficiently over the proposed sampling scheme from its spherical harmonic coefficients. We later show that the proposed sampling scheme and SHTs~(both forward and inverse) are also applicable for band-limited spin functions $\fs\in{}_s\lsphL{L}$.

\subsection{Proposed Sampling Scheme}\label{sec:sampling_one}

We propose an iso-latitude sampling of the sphere. Define the indexed vector $\bv{\theta}$ as
\begin{align}
\bv\theta \dfn [\theta_0,\, \theta_1,\,\hdots,\,\theta_{L-1}]^T,
\end{align}
which consists of $L$ points along $\theta$. Also define $\bv\theta^{m} \dfn [\theta_{|m|}, \theta_{|m|+1},\hdots,\theta_{L-1}]^T$ as a vector of $L-|m|$ arbitrary points along $\theta$ for $ |m| < L$. We shortly present the location of these sample points.
For discretization along $\phi$, we consider $2k+1$ equally spaced sampling points along $\phi$ for each $\theta_k \in \bv\theta $.
Define $\bv\phi^{k}$ be a vector of $2k+1$ equally spaced sampling points along $\phi$ in the ring placed at $\theta_k$, given by
\begin{equation}
\bv\phi^k \dfn [0,\, \Delta_k,\,2\Delta_k,\, \hdots,\, (2k)\Delta_k],\quad \Delta_k = \frac{2\pi}{2k+1}.
\end{equation}
In this way, we will have $L$ iso-latitude rings of sampling points along $\phi$ for each sampling point $\theta_k$ along $\theta$, where the number of points, $2k+1$, in each ring depends on the location of the ring along latitude. We note that the number of samples in the proposed sampling scheme attains the optimal spatial dimensionality, that is,
\begin{equation}\label{Eq:number_samples_sum}
\sum_{k=0}^{L-1}(2k+1) = L^2 = \nps,
\end{equation}
which also represents the number of degrees of freedom in harmonic space for a signal band-limited at $L$. We first develop the forward and inverse SHTs and later provide details about the location of the samples $\theta_k$ in the vector $\bv{\theta}$.

\subsection{Forward Spherical Harmonic Transform -- Formulation}

We develop the forward SHT to compute the spherical harmonic coefficients $\shc{f}{\ell}{m}$ of a band-limited signal $f$ sampled over the $\nps$ samples of our sampling scheme. First, we first develop the necessary mathematical formulation and later we present the philosophy of our approach and develop the forward SHT. For order $|m| < L$, define a vector
\begin{equation}
\mathbf{g}_m \equiv G_m(\bv\theta^{m}) \dfn [G_m(\theta_{|m|}),\,G_m(\theta_{{|m|}+1}),\, G_m(\theta_{L-1})]^T,
\end{equation}
with
\begin{align}\label{Eq:gm_integral}
G_m(\theta_k) &\dfn \intphi f(\theta_k,\phi) e^{-im\phi} d\phi \nonumber \\
            &= 2\pi \sum_{\ell=m}^{L-1} \shc{f}{\ell}{m}  \plms_\ell^m(\theta_k),
\end{align}
for each $\theta_k\in \bv{\theta}$, where $\plms_\ell^m(\theta) \dfn Y_\ell^m(\theta,0)$ denotes scaled associated Legendre functions~(see \appref{App:maths} for the definition of associated Legendre functions and spherical harmonic $Y_\ell^m$)

By defining a matrix $\mathbf{P}_m$ as
\begin{equation}\label{Eq:Y_matrix}
\mathbf{P}_m \dfn  2\pi\begin{small}\setlength{\arraycolsep}{1mm}
\begin{pmatrix}
   \plms_{|m|}^m(\theta_{|m|}) & \plms_{{|m|}+1}^m(\theta_{|m|}) & \cdots & \plms_{L-1}^m(\theta_{|m|}) \\[2mm]
   \plms_{|m|}^m(\theta_{{|m|}+1}) & \plms_{{|m|}+1}^m(\theta_{{|m|}+1}) & \cdots & \plms_{L-1}^m(\theta_{{|m|}+1}) \\
   \vdots  & \vdots  & \ddots & \vdots  \\
   \plms_{|m|}^m(\theta_{L-1}) & \plms_{{|m|}+1}^m(\theta_{L-1}) & \cdots & \plms_{L-1}^m(\theta_{L-1})\\
  \end{pmatrix}\end{small},
\end{equation}
and a vector $\mathbf{f}_m$ containing spherical harmonic coefficients of order $|m| < L$ given by
\begin{equation}
\mathbf{f}_m  = \big[ \shc{f}{{|m|}}{m},\,  \shc{f}{{|m|}+1}{m},\,\hdots,\, \shc{f}{L-1}{m}\big]^T,
\end{equation}
we can write $\mathbf{g}_m$ as follows
\begin{equation}\label{Eq:gtof_inverse}
\mathbf{g}_m  = \mathbf{P}_m \mathbf{f}_m,
\end{equation}
where the vector $\mathbf{g}_m$ contains the values $G_m(\theta_k)$ for $\theta_k\in \bv\theta^{m}$. It is possible to recover $\mathbf{f}_m$ by inverting this system, which is elaborated shortly.

\begin{remark}\label{remark:transform}
Using the formulation in \eqref{Eq:gtof_inverse}, $L-{|m|}$ number of spherical harmonic coefficients of order $m$~(or $-m$) contained in the vector $\mathbf{f}_m$~(or $\mathbf{f}_{-m}$) can be determined by first computing $\mathbf{g}_m$~(or $\mathbf{g}_{-m}$) over $L-{|m|}$ number of samples or equivalently by evaluating $G_m(\theta)$~(or $G_{-m}(\theta)$ ) for all $\theta_k \in \bv \theta^{m}$ and the matrix $\mathbf{P}_m$ in \eqref{Eq:Y_matrix} and then solving \eqref{Eq:gtof_inverse}. We note that $\mathbf{P}_m = (-1)^m\mathbf{P}_{-m}$~(using \eqref{eq:yellm:conj}), $\mathbf{P}_m$ is not dependent on the signal and $\bv\theta^{m}$ must be chosen such that $\mathbf{P}_m$ can be inverted accurately, which enables the accurate computation of $\mathbf{f}_m$~(or $\mathbf{f}_{-m}$).
\end{remark}

\subsection{Forward Spherical Harmonic Transform -- Philosophy}
Following Remark~\ref{remark:transform}, we need to compute $G_m(\theta)$ formulated in \eqref{Eq:gm_integral} for all $\theta_k \in \bv \theta^{m}$ and for given $|m|< L$. Using an FFT, $G_m(\theta)$ for each $m$ can be computed exactly by evaluating the integral as a summation, provided the following two conditions are satisfied:

\begin{itemize}
\item $\intphi f(\theta_k,\phi) e^{-im'\phi} d\phi = 0$, for all $|m'|>|m|$ and for all $\theta_k\in \bv \theta^{m}$, which ensures that the univariate signal $f(\theta_k,\phi)$ along $\phi$ for each $\theta_k$ is band-limited at $|m|+1$, given the complex exponentials $e^{im'\phi}$ as basis functions along $\phi$, and
\item there are at least $2|m|+1$ sample points in each $\phi$-ring placed for all $\theta_k \in \bv \theta^{m}$.
\end{itemize}
These conditions and the following remark form the foundation of our proposed method to compute the forward SHT.

\begin{remark}\label{remark:bandlimit}
For a band-limited signal on the sphere given by \eqref{Eq:f_expansion_bandlimited}, we have contributions from the complex exponentials $e^{im'\phi}$ for $0 \leq |m'| \leq (L-1)$; thus we require $2L-1$ samples in \emph{each} of the rings placed at $\theta_k \in \bv \theta^{m}$ in order to compute $G_m(\theta_k)$ correctly regardless of the choice of $m$. However, if the spherical harmonic coefficients of all degrees~(and orders) greater than $|m|$ are known for some $|m|< L$, their contributions can be removed. Such a removal makes the signal band-limited along $\phi$ with respect to the contributions of the complex exponentials $e^{i|m'|\phi}$ for $0 \leq |m'| \leq |m|$.  Consequently, $G_m(\theta_k)$ can be computed correctly by taking an FFT over \emph{only} $2|m|+1$ samples (instead of $2L-1$ samples) in a ring placed at some $\theta_k \in \bv \theta^{m}$. We elaborate this philosophy below.
\end{remark}

\begin{figure}[t]
    \centering
    \includegraphics[scale=0.4]{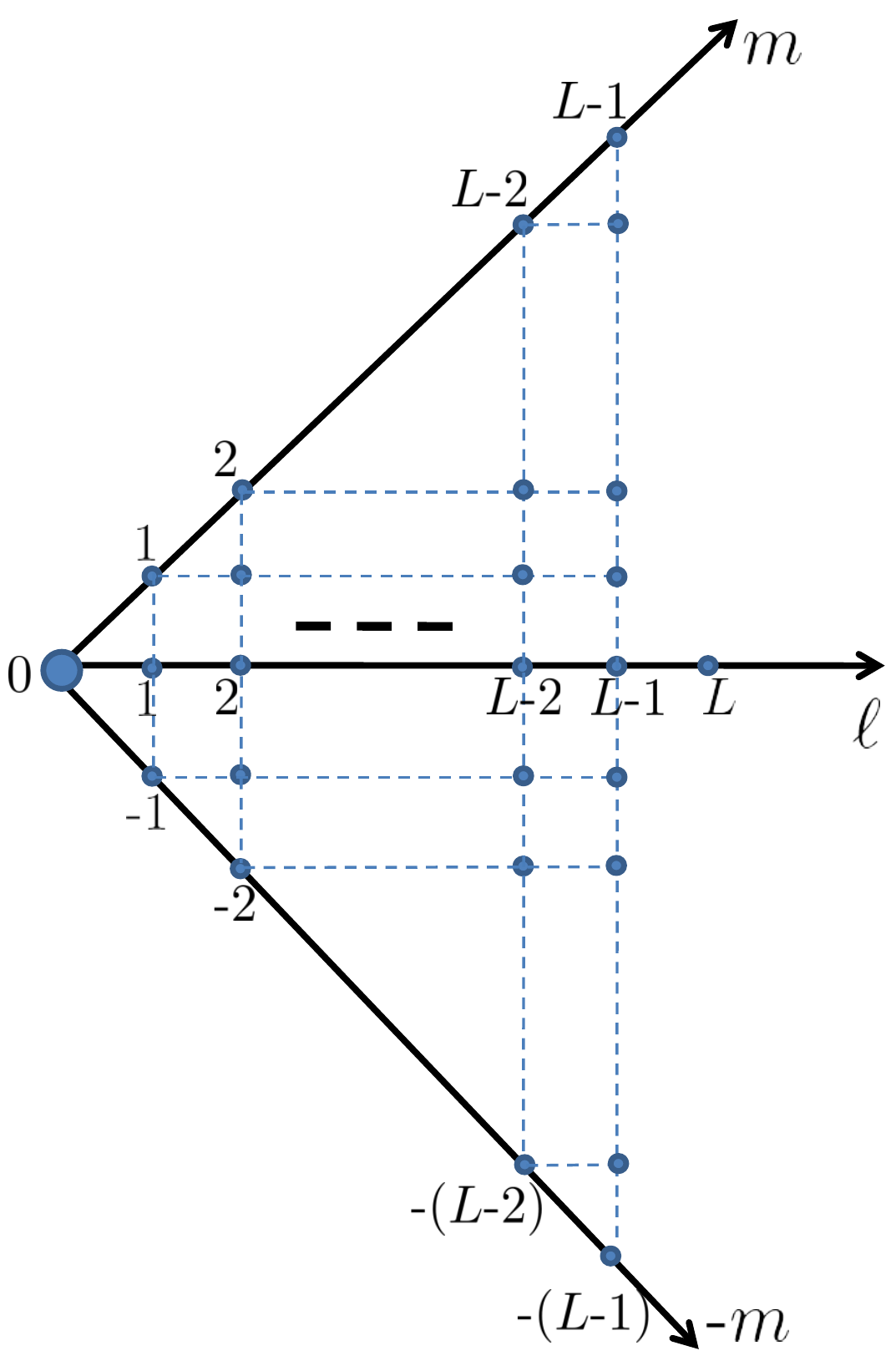}
    \caption{The graphical representation of the spectral domain~(formed by spherical harmonic coefficients) of a signal band-limited at $L$. For each $-L<m<L$, there are $L-|m|$ spherical harmonic coefficients.}
    \label{fig:spectral_domain_representation}
\end{figure}

\figref{fig:spectral_domain_representation} shows the graphical representation of the spectral domain of a signal band-limited at $L$. There is one $L-1$ order spherical harmonic coefficient $\shc{f}{L-1}{L-1}$ and one $-(L-1)$ order spherical harmonic coefficient $\shc{f}{-(L-1)}{-(L-1)}$, which can be determined using \eqref{Eq:gtof_inverse} by first computing $\mathbf{g}_{L-1}  = G_{L-1}(\theta_{L-1})$ and $\mathbf{g}_{-(L-1)}  = G_{-(L-1)}(\theta_{L-1})$ using an FFT over only one ring of $2L-1$ samples along $\phi$ placed at $\theta_{L-1}$. It should be noted that an FFT over $2L-1$ samples, in fact, computes $G_m(\theta_k)$ correctly for all orders $ -L  < m <  L $. Once $\shc{f}{L-1}{L-1}$ is computed, the signal at the other sample positions~(for all $\theta_k \in \bv\theta\backslash\bv\theta^{L-1}$ and associated
samples along $\phi$ in each ring) can be updated as
\begin{equation}\label{Eq:signal_update}
f(\theta,\phi) \leftarrow f(\theta,\phi)-\tilf_{L-1}(\theta,\phi)
\end{equation}
where
\begin{align}\label{Eq:signal_subset}
\tilf_m(\theta,\phi) &= \sum_{{\ell}=m}^{L-1} \left(\shc{f}{\ell}{m}  Y_{\ell}^m(\theta,\phi) + \shc{f}{\ell}{-m}  Y_{\ell}^{-m}(\theta,\phi) \right)\nonumber \\
&= \sum_{{\ell}=m}^{L-1} \left( \shc{f}{\ell}{m}  \plms_{\ell}^m(\theta)e^{im\phi} + \shc{f}{\ell}{-m}  \plms_{\ell}^{-m}(\theta)e^{-im\phi} \right) \nonumber
\\
&= \frac{1}{2\pi}\left(e^{im\phi} G_m(\theta) + e^{-im\phi} G_{-m}(\theta)\right)
\end{align}
denotes the part of the signal $f(\theta,\phi)$ composed of contribution of spherical harmonics of order $m$ and $-m$ and all degrees $m \leq \ell\leq (L-1)$. Once the signal is updated at other sample positions as given in \eqref{Eq:signal_update}, there is no contribution of spherical harmonics of orders $-(L-1)$ and $(L-1)$ or equivalently there is no contribution of complex exponentials $e^{i(L-1)\phi}$ and $e^{-i(L-1)\phi}$ in the signal. Thus, following the conditions stated earlier in this subsection, we need $2L-3$~(instead of $2L-1$) samples along $\phi$ in the rings placed at all $\theta_k \in \bv\theta\backslash\bv\theta^{L-1}$. We only require $2L-3$ samples along the $\phi$-ring placed at $\theta_{L-2}$ to determine the spherical harmonic coefficients of order $L-2$ and $-(L-2)$, which once computed can be used to update the signal as
\begin{equation}\label{Eq:signal_update_2}
f(\theta,\phi)   \leftarrow f(\theta,\phi)-\tilf_{L-2}(\theta,\phi),
\end{equation}
at other sample positions for all $\theta_k \in \bv\theta\backslash\bv\theta^{L-2}$ and samples in the associated
samples along $\phi$ in each ring. Proceeding in this similar manner, we note that the forward SHT can be computed using $2k+1$ number of samples along each ring $\bv\phi^{k}$ placed at $\theta_k, \, k\in[0,1,\hdots,L-1]$. Using the philosophy mentioned above, we summarize the proposed forward SHT in the form of the procedure below.

\begin{algorithm}
\caption{Forward Spherical Harmonic Transform}\label{Procedure:FSHT}
\begin{algorithmic}[1]
\Require $\shc{f}{\ell}{m}$,\quad $\forall$  $0\leq \ell \leq (L-1),\, |m| \leq \ell$, given $f(\theta,\phi)$
\Procedure{Forward SHT}{$f(\theta,\phi)$}
\For{$m =L-1,L-2,\hdots,0$}
\State compute $\mathbf{g}_m$ and $\mathbf{g}_{-m}$ by evaluating $G_m(\theta)$ and
\NoNumber{$G_{-m}(\theta)$ for all $\theta_k \in \bv\theta^m$ by taking $(2m+1)$ point}
\NoNumber{FFT along each $\phi$-ring}
\vspace{1.5mm}
\State evaluate $\mathbf{P}_m$ and $\mathbf{P}_{-m}$, using $\mathbf{P}_{-m} = (-1)^m \mathbf{P}_{m}$
\vspace{1.5mm}
\State compute $\mathbf{f}_m$ and $\mathbf{f}_{-m}$ by inverting \eqref{Eq:gtof_inverse}
\vspace{1.5mm}
\State determine $\tilf_m(\theta,\phi)$ for all $\theta_k \in \bv\theta \backslash \bv\theta^m $ and
\NoNumber{all associated sampling points along $\phi$}
\vspace{1.5mm}
\State update $f(\theta,\phi) \leftarrow f(\theta,\phi)-\tilf_m(\theta,\phi)$ for all $\theta_k \in $
\NoNumber{$\bv\theta \backslash \bv\theta^m $ and all associated sampling points along $\phi$}
\EndFor
\State \textbf{return} $\shc{f}{\ell}{m}$
\EndProcedure%
\end{algorithmic}
\end{algorithm}

\subsection{Inverse Spherical Harmonic Transform}
The inverse SHT computes the signal from its spherical harmonic coefficients. Using the separation of variables technique~(also adopted in~\cite{Driscoll:1994,Kostelec:2008,Wiaux:2005:2,McEwen:2011}), changing the order of summation in \eqref{Eq:f_expansion} and using \eqref{Eq:gm_integral} and \eqref{Eq:signal_subset}, we write the inverse SHT as follows
\begin{align}\label{Eq:f_inverse_SHT}
f(\theta,\phi) &= \frac{1}{2\pi}\sum_{m=-(L-1)}^{L-1} e^{im\phi} G_m(\theta)\nonumber
\\
&=  \sum_{m=0}^{L-1} \tilf_m(\theta,\phi),
\end{align}
where $\theta$ and $\phi$ are the sample points belonging to the proposed sampling scheme. The inverse SHT can be computed by the following procedure.

\begin{algorithm}
\caption{Inverse Spherical Harmonic Transform}\label{Procedure:ISHT}
\begin{algorithmic}[1]
\Require $f(\theta,\phi)$, given $\shc{f}{\ell}{m}$,\quad $\forall$  $0\leq \ell \leq (L-1),\, |m| \leq \ell$
\Procedure{Inverse SHT}{$\shc{f}{\ell}{m}$}
\\
initialization\; $f(\theta,\phi)=0$
\For{$m =0,1,\hdots,L-1$}
\State evaluate $\plms_\ell^m(\theta_k)$ and $\plms_\ell^{-m}(\theta_k)$ for all $\theta_k\in\bv{\theta}$ and
\NoNumber{$m\leq \ell \leq (L-1)$}
\vspace{1.5mm}
\State compute $G_m(\theta_k)$ and $G_{-m}(\theta_k)$ using \eqref{Eq:gm_integral} for all
\NoNumber{$\theta_k\in\bv{\theta}$}
\vspace{1.5mm}
\State compute $\tilf_m(\theta,\phi)$ using \eqref{Eq:signal_subset} for all sampling
\NoNumber{points}
\vspace{1.5mm}
\State update $f(\theta,\phi) \leftarrow f(\theta,\phi)+\tilf_m(\theta,\phi)$ for all
\NoNumber{sampling points}
\EndFor
\State \textbf{return} $f(\theta,\phi)$
\EndProcedure%
\end{algorithmic}
\end{algorithm}

\begin{figure*}[t]
    \centering
    \hspace{-4mm}
    \subfloat[]{
        \includegraphics[width=0.31\textwidth]{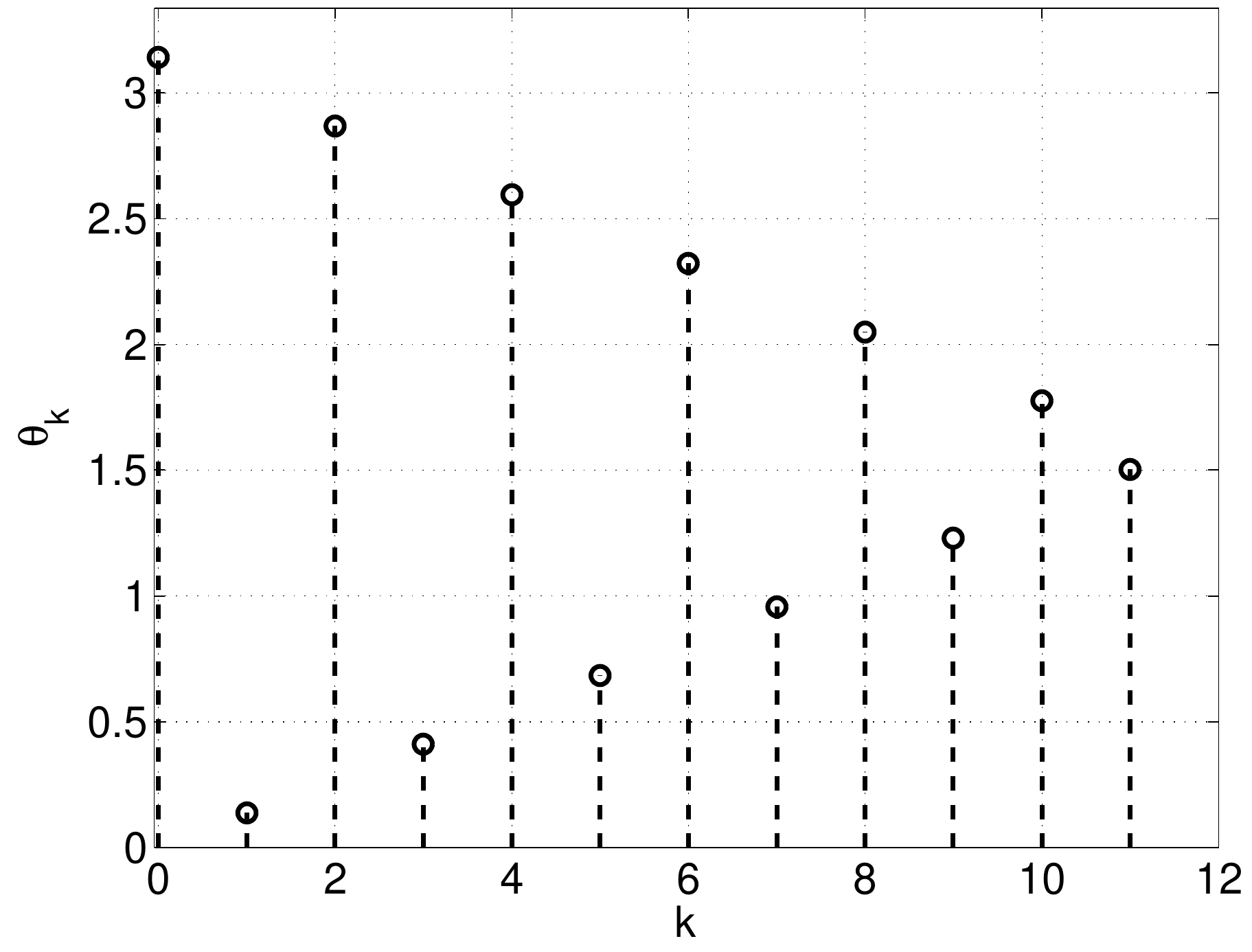}}\hfil
   \hspace{-3mm}
    \subfloat[]{
        \includegraphics[width=0.36\textwidth]{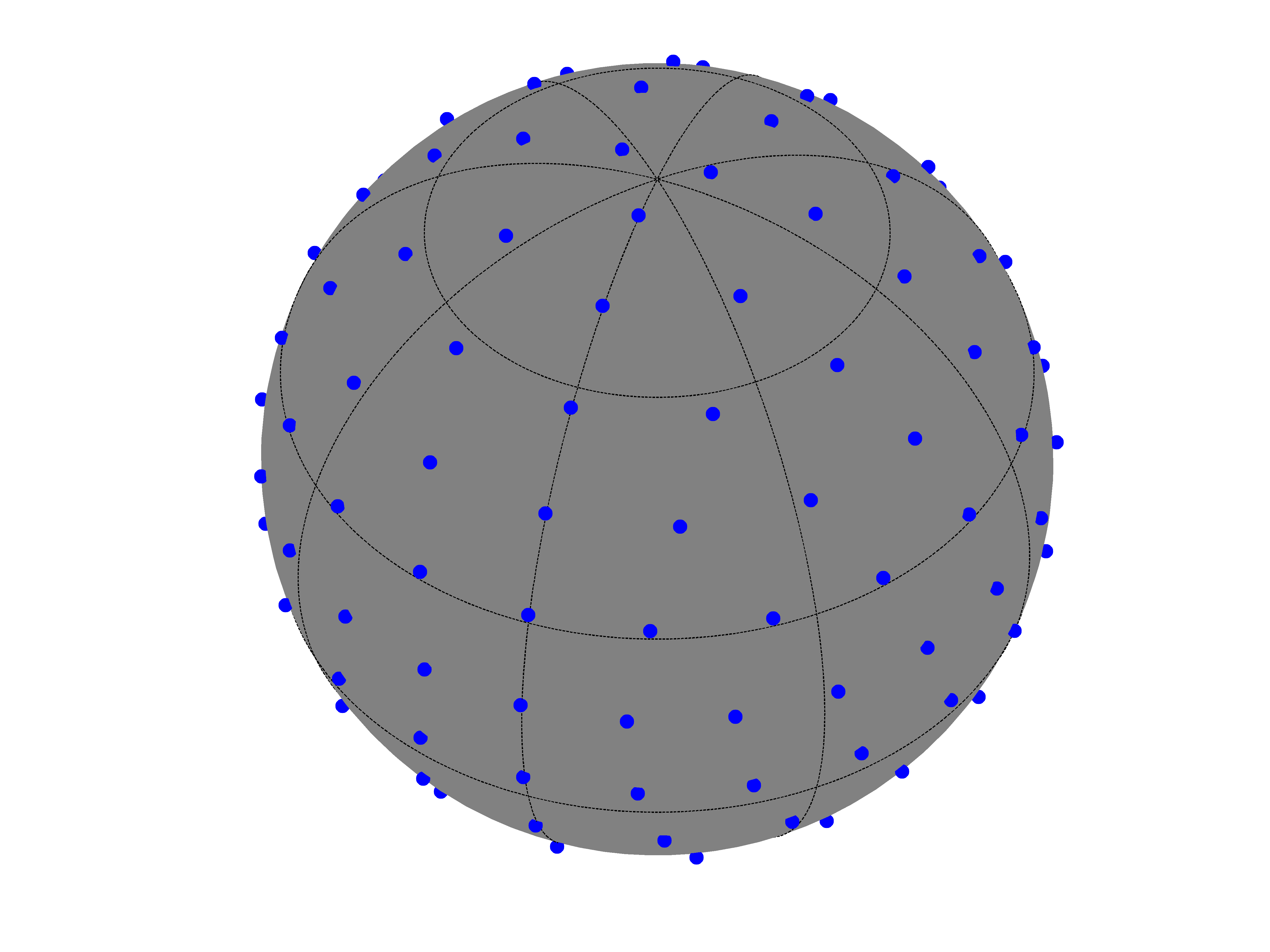}}\hfil
   \hspace{-13mm}
    \subfloat[]{
        \includegraphics[width=0.36\textwidth]{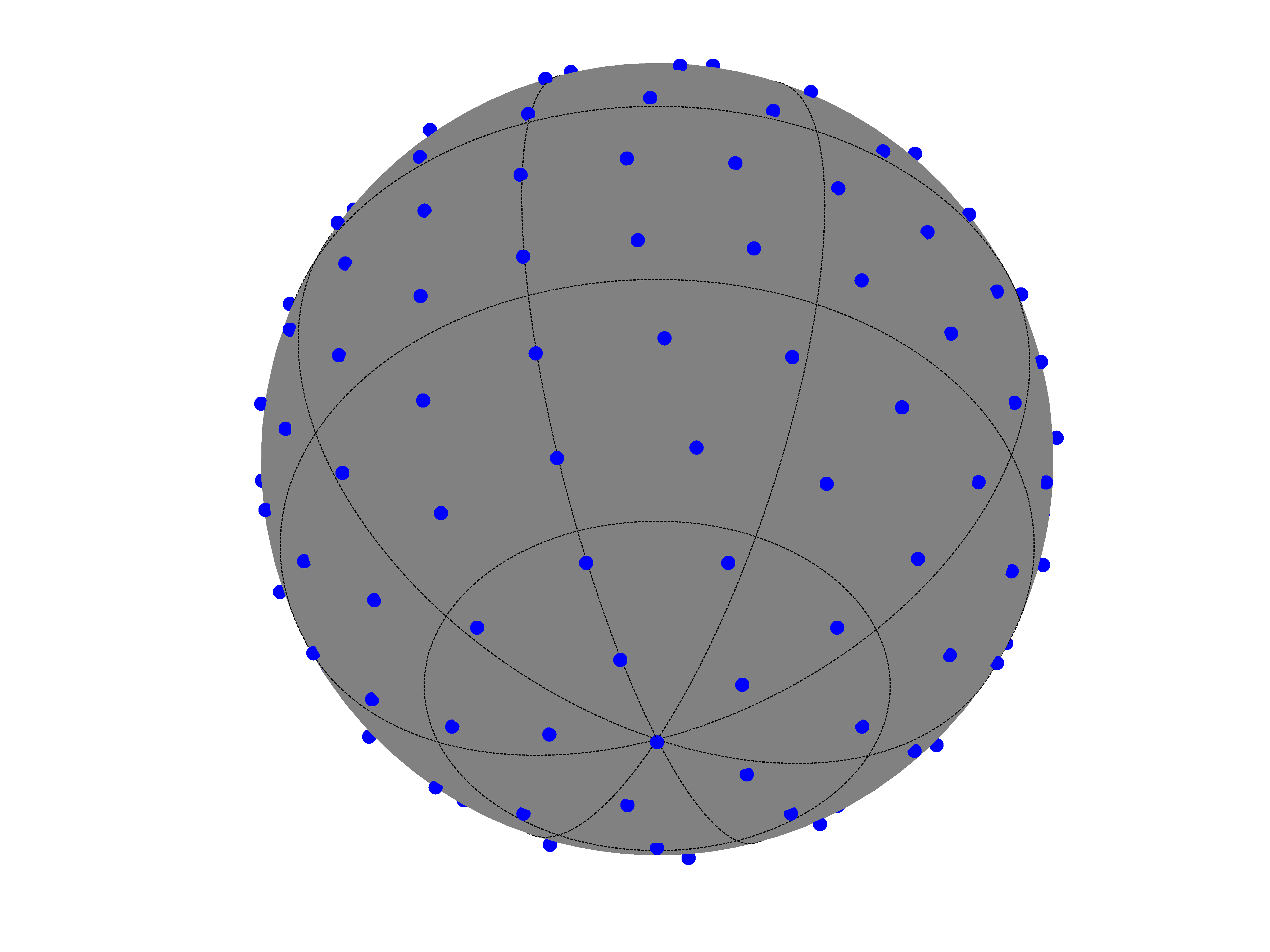}}\\[-3mm]

    \vspace{2mm}

    \caption{The sampling scheme on the sphere given in \secref{sec:sampling_one} for the representation of the signal band-limited at $L=12$. (a) The samples along co-latitude $\theta_k$ in a vector $\bv\theta$ versus the index $k$ given in \eqref{Eq:sampling_latitude}. The samples on the sphere are shown with a view from (a) North Pole and (b) South Pole.}
    \label{fig:sample_positions}
\end{figure*}
\begin{figure*}[t]
    \centering
    \hspace{-4mm}
    \subfloat[]{
        \includegraphics[width=0.32\textwidth]{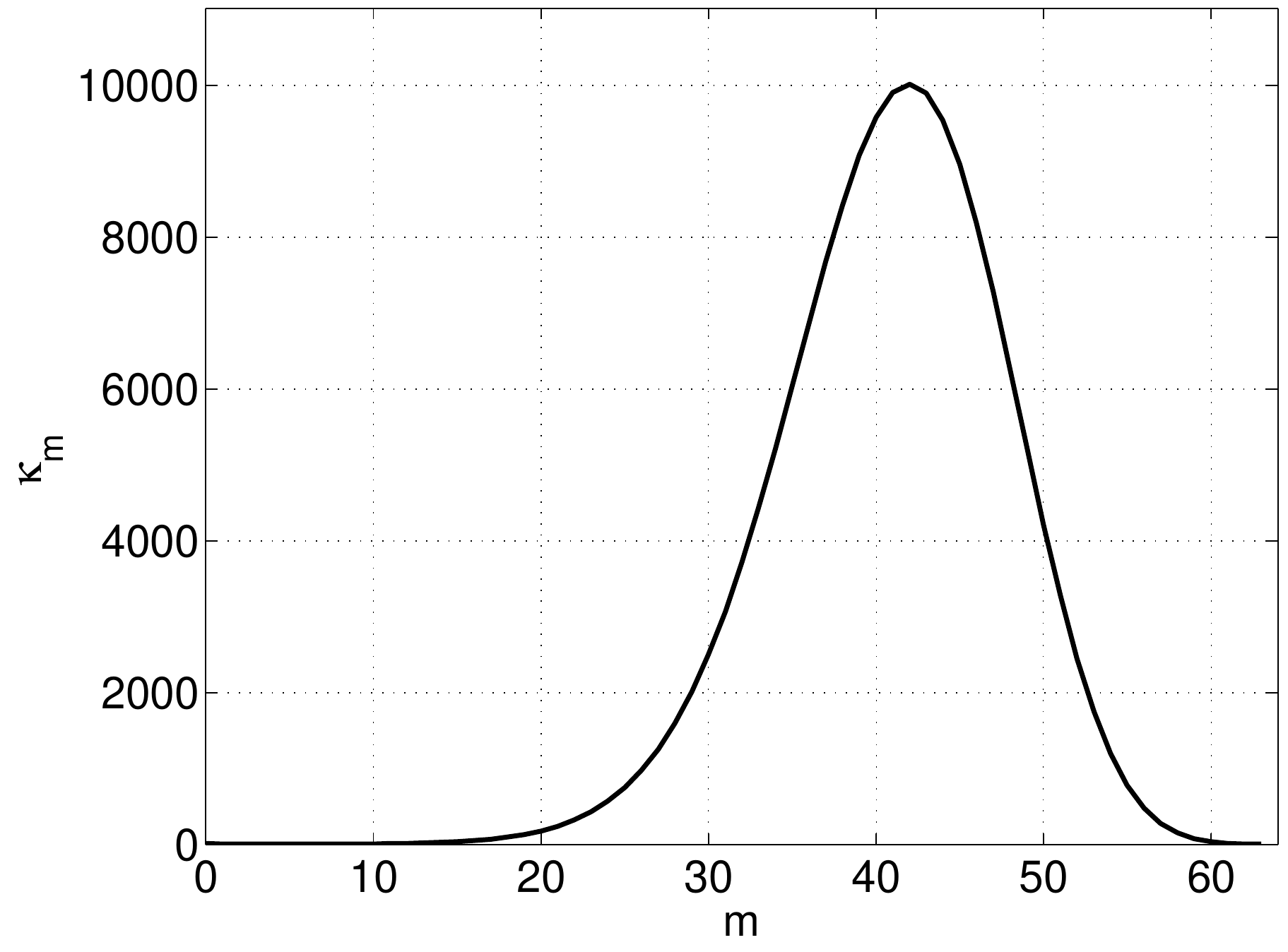}}\hfil
   \hspace{-1mm}
    \subfloat[]{
        \includegraphics[width=0.32\textwidth]{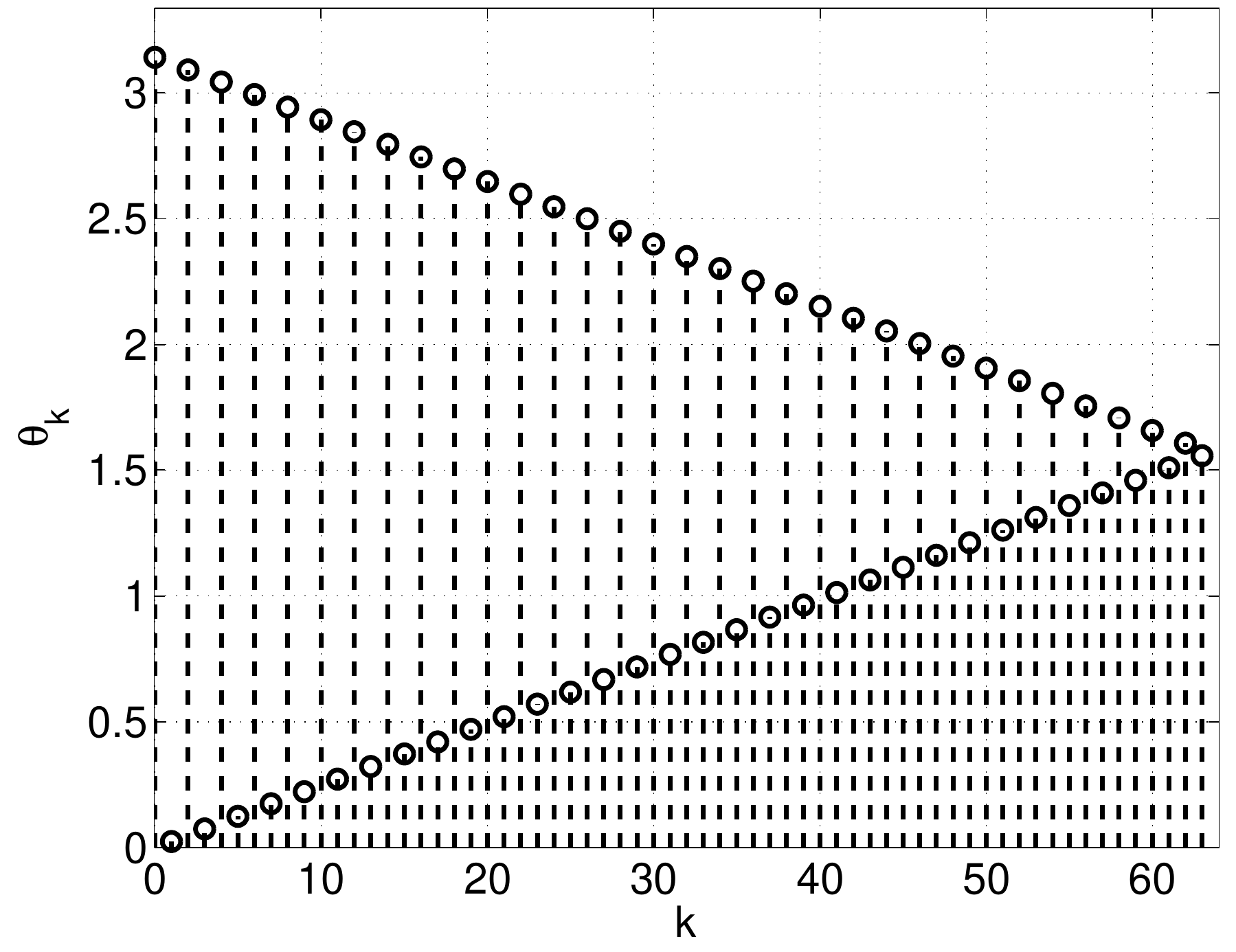}}\hfil
   \hspace{-1mm}
    \subfloat[]{
        \includegraphics[width=0.32\textwidth]{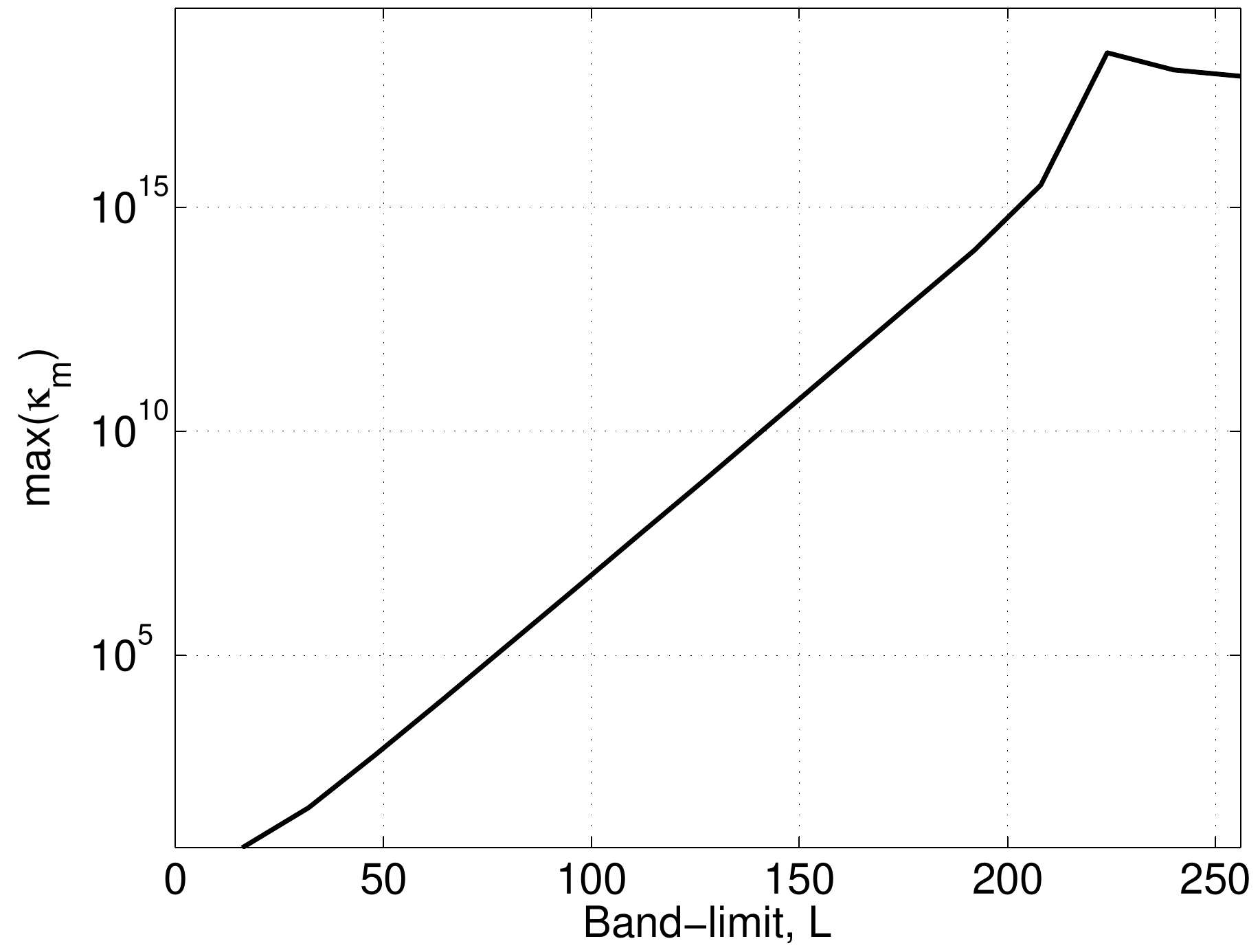}}\hfil
    \caption{(a) The condition number $\kappa_m$ of the matrix $\mathbf{P}^m$ for $L=64$ and for different values of $0 \leq m<L$, where the matrix $\mathbf{P}^m$ is constructed with the sample positions  along co-latitude in (b) a vector $\bv{\theta}$ given in \eqref{Eq:sampling_latitude}. (c) The maximum of the condition number, $\max(\kappa_m),  \, 0\leq m < L$ for different band-limit $16 \leq L \leq 512$. Note that the maximum condition number $\max(\kappa_m)$ quickly grows to large values as the band-limit $L$ increases. }
    \label{fig:cond_number}
\end{figure*}

\begin{figure*}[t]
    \centering
    \hspace{-4mm}
    \subfloat[]{
        \includegraphics[width=0.32\textwidth]{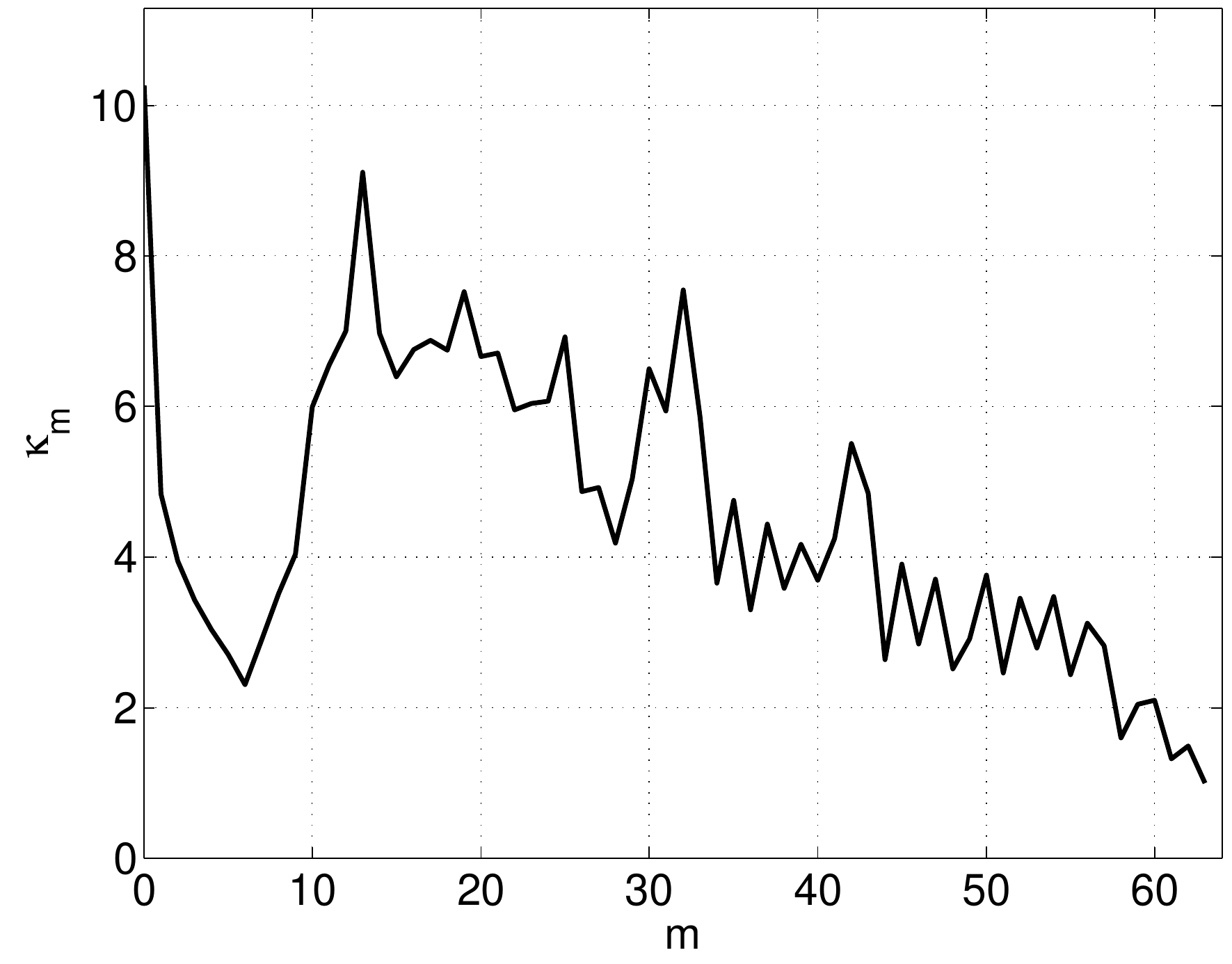}}\hfil
    \subfloat[]{
        \includegraphics[width=0.32\textwidth]{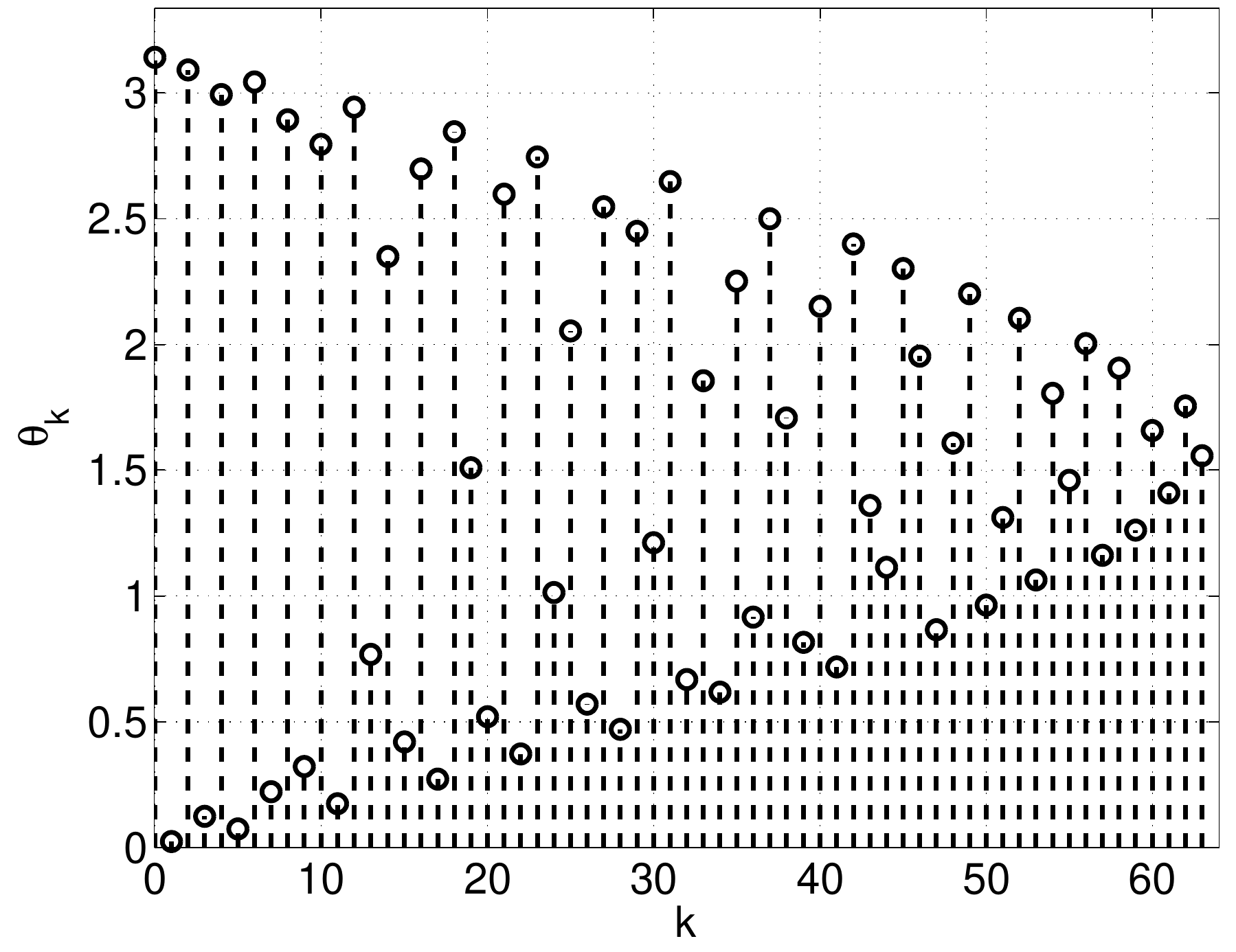}}\hfil
    \subfloat[]{
        \includegraphics[width=0.32\textwidth]{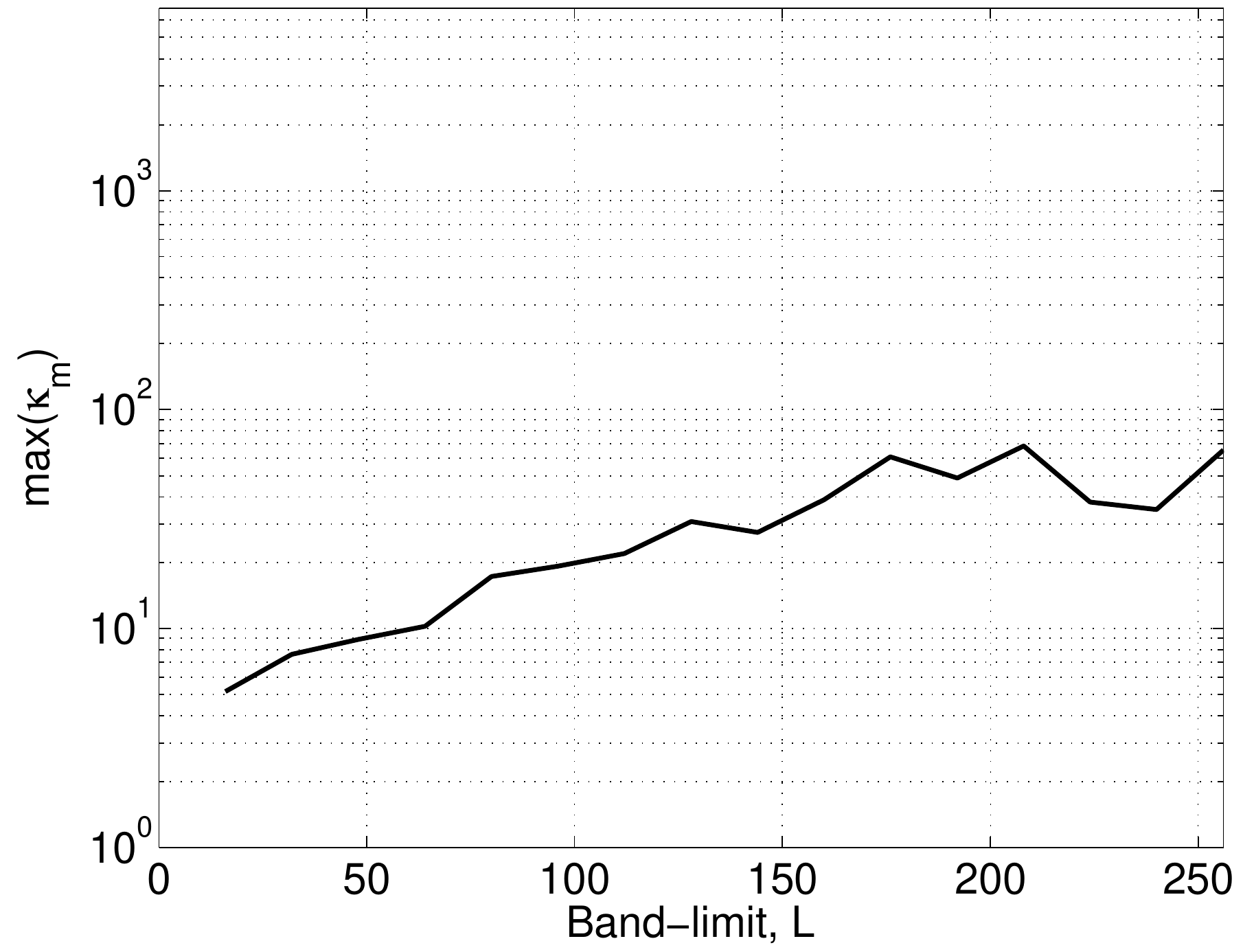}}\hfil
    \caption{(a) The condition number $\kappa_m$ of the matrix $\mathbf{P}^m$ for $L=64$ and for different values of $0 \leq m<L$, where the matrix $\mathbf{P}^m$ is constructed with the optimal sample positions along co-latitude in (b) a vector $\bv{\theta}$ obtained using the condition number minimization method presented in \secref{sec:placement_samples}. (c) The maximum of the condition number, $\max(\kappa_m),  \, 0\leq m < L$ for different band-limit $16 \leq L \leq 512$. Note that the maximum condition number $\max(\kappa_m)$ is significantly smaller for the case of optimal sampling, resulting in an accurate implementation of the proposed forward SHT method.}
    \label{fig:cond_number_optimal}
\end{figure*}

\subsection{Placement of Samples along Co-latitude}\label{sec:placement_samples}

For the representation of band-limited signals with band-limit $L$ we yet need to choose the location of $L$ samples, $\theta_k\in\bv{\theta},\, k\in[0,1,\hdots,L-1]$,  along co-latitude for the sampling scheme presented in \secref{sec:sampling_one}, where a ring of $2k+1$ equiangular sample points along $\phi$ is placed at each $\theta_k$.  The simplest choice is to use the equiangular set of samples given by
\begin{equation}\label{Eq:equiangular_set}
\Theta = \bigg\{\frac{\pi(2t+1)}{2L-1} \bigg\}, \quad t=0,1,\hdots,L-1,
\end{equation}
for the placement of the rings, where rings are placed such that the rings with greater numbers of samples are placed nearer to the equator $\theta=\pi/2$.
In this setting, the ring of $1$ sample is placed at $\theta_0=\pi$, the ring of $3$ samples is placed at $\theta_1 = \frac{\pi}{2L-1}$ and so on. For such a placement of samples, the vector $\bv{\theta}$ for band-limit $L$ is given by
\begin{equation}\label{Eq:sampling_latitude}
\begin{small}
\bv\theta \dfn \bigg[\pi,\, \frac{\pi}{2L-1},\, \frac{\pi(2L-3)}{2L-1},\, \frac{3\pi}{2L-1},\,   \hdots,\,    \frac{\pi(2\lfloor\frac{L-1}{2}\rfloor+1)}{2L-1} \bigg]^T.
\end{small}
\end{equation}
As an example, the samples on the sphere for this scheme are shown in \figref{fig:sample_positions} for $L=12$.

We note that the proposed forward SHT requires the samples in the vector $\theta$ to be chosen such that the matrix $\mathbf{P}^m$ becomes invertible so that the system in \eqref{Eq:gtof_inverse} can be inverted accurately. The sampling along the co-latitude as given in \eqref{Eq:sampling_latitude}, although an attractive choice, may not be appropriate as the matrix $\mathbf{P}^m$ may become ill-conditioned.
For example, the condition number~(ratio of the largest eigenvalue to the smallest eigenvalue), denoted by $\kappa_m$, of the matrix $\mathbf{P}^m$, constructed with the sample positions $\bv{\theta}$ given in \eqref{Eq:sampling_latitude}, for $L=64$ and for different values of $0 \leq m<L$ is plotted in \figref{fig:cond_number}(a) and the samples in a vector $\bv{\theta}$ are shown in \figref{fig:cond_number}(b). Furthermore, the maximum of the condition number, denoted by $\max(\kappa_m)$ over  $0\leq m < L$ for different values of band-limit $L$ is plotted in \figref{fig:cond_number}(c), where it can be observed that at least one matrix $\mathbf{P}^m$ for $0\leq m < L$ becomes more ill-conditioned for larger band-limit $L$.

In order to address this issue, we propose the following recipe to determine the optimal ordering of samples in a vector $\bv\theta$, which we refer to as the condition number minimization method. For a given $L$ and samples in $\theta$ given by \eqref{Eq:equiangular_set}, the vector $\bv\theta$ is constructed as follows:

\begin{itemize}
\item Choose $\theta_{L-1}=\frac{\pi(2\lfloor\frac{L-1}{2}\rfloor+1)}{2L-1}$ farthest from the poles, which is a natural choice for the ring of $2L-1$ samples along $\phi$.
\item For each $m=L-2,\,L-3,\,\hdots,\,0$, choose $\theta_{m}$ from the set $\Theta$, given in \eqref{Eq:equiangular_set}, which minimizes the condition number of the matrix $\mathbf{P}^m$.
\end{itemize}
Such a placement of samples along co-latitude ensures the robust inversion of the system given by \eqref{Eq:gtof_inverse}, thus resulting in an accurate computation of spherical harmonic coefficients by the proposed forward SHT. As an illustration, we again plot the condition number $\kappa_m$ of the matrix $\mathbf{P}^m$ obtained using the optimal sample positions for $L=64$ in \figref{fig:cond_number_optimal}(a), the optimal sample positions in \figref{fig:cond_number_optimal}(b) and the maximum condition number $\max(\kappa_m$) for different band-limits $L$ in \figref{fig:cond_number_optimal}(c). In comparison to the plots in \figref{fig:cond_number}, the condition number is significantly smaller for the case of optimal sampling, which leads to an accurate implementation of the proposed forward SHT. Furthermore, when computing multiple harmonic transforms for different $L$, the optimal position of
samples in a vector $\bv\theta$ needs to be computed once \emph{only} for each $L$. The vector $\bv\theta$ can be stored in a double precision with the storage requirement of only $30.4$KB for $L=4096$, for example, and approximately $60$MB for all band-limits \mbox{$L<1024$}. Moreover, we highlight that this storage determines the placement of iso-latitude rings on the sphere and is required to be known for either forward or inverse transform. Once the sample positions are known, we do not require any further precomputation for the computation of SHTs. We discuss the computational complexity of the proposed transforms later in the paper.

\subsection{Alternative Placement of Samples along Co-latitude}
The equiangular samples in the set $\Theta$ are placed along co-latitude $\theta$ according to a uniform measure $d\theta$. Alternatively, the samples along co-latitude can be placed according to different measures. For example, in the context of compressive sensing, it has been proved in \cite{Burq:2012} that a sparse~(in spectral domain) band-limited signal can be recovered
from fewer measurements if samples are drawn from the measure $|\tan\theta|^{1/3}d\theta d\phi$, compared to sampling with respect to
the uniform measure $d\theta d\phi$, which in turn has been shown by \cite{Rauhat:2011} to require fewer samples than sampling with respect to
the measure $\sin\theta d\theta$.

We compare the equiangular placement of rings with the placement of rings according to the measures $\sin\theta d\theta$ and $|\tan\theta|^{1/3}d\theta d\phi$. Define $\Theta^1 = \big\{\Theta^1_t\big\}$ and $\Theta^2 = \big\{\Theta^2_t\big\}$ for $t=0,1,\hdots,L-1$ as sets of $L$ samples along co-latitude, where samples are placed according to the measures $\sin\theta d\theta$ and $|\tan\theta|^{1/3}d\theta$, respectively. The sets $\Theta^1$ and $\Theta^2$ are constructed by choosing $\Theta^1_0 = \Theta^2_0 = \pi$~(South Pole) and using the following relation between the consecutive samples
\begin{align*}
\int_{\Theta^1_{t}}^{\Theta^1_{t-1}} \sin\theta d\theta &= \frac{1}{L}\int_{0}^{\pi} \sin\theta d\theta = \frac{2}{L}, \\
\int_{\Theta^1_{t}}^{\Theta^1_{t-1}} |\tan\theta|^{1/3} d\theta &= \frac{1}{L}\int_{0}^{\pi} |\tan\theta|^{1/3} d\theta   = \frac{2\pi}{L\sqrt{3}},
\end{align*}
for  $t=1,2,\hdots,L-1$. Using the sets $\Theta^1$ and $\Theta^2$, we construct the indexed vectors $\bv\theta^1$ and $\bv\theta^2$, respectively, similar to $\bv\theta$ in \eqref{Eq:sampling_latitude} defined for the set $\Theta$ in \eqref{Eq:equiangular_set}, to determine the sample positions of iso-latitude rings such that the ring with samples is placed nearer to the equator.

\begin{figure*}[t]
    \centering
    \hspace{-4mm}
    \subfloat[]{
        \includegraphics[width=0.32\textwidth]{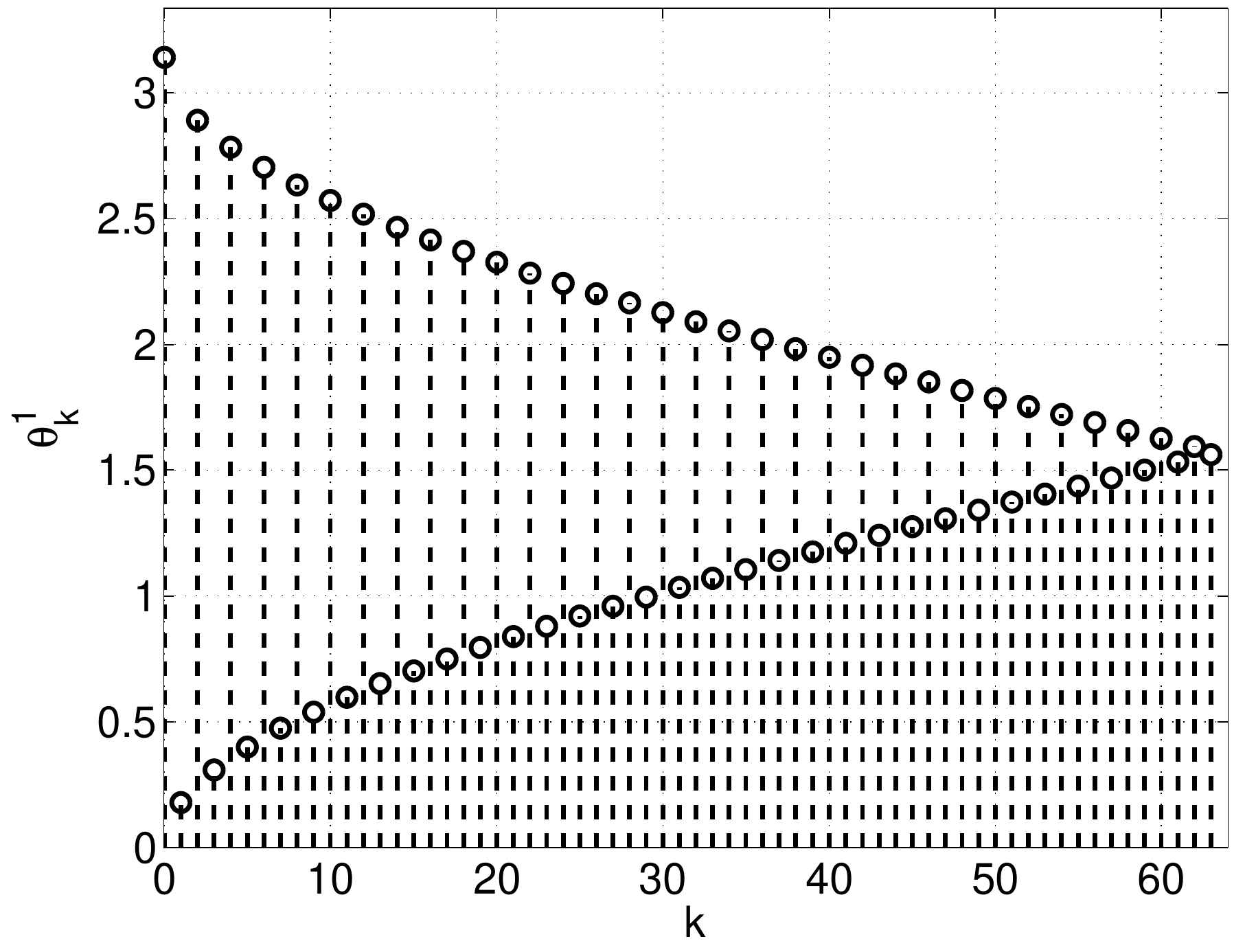}}\hfil
    \subfloat[]{
        \includegraphics[width=0.32\textwidth]{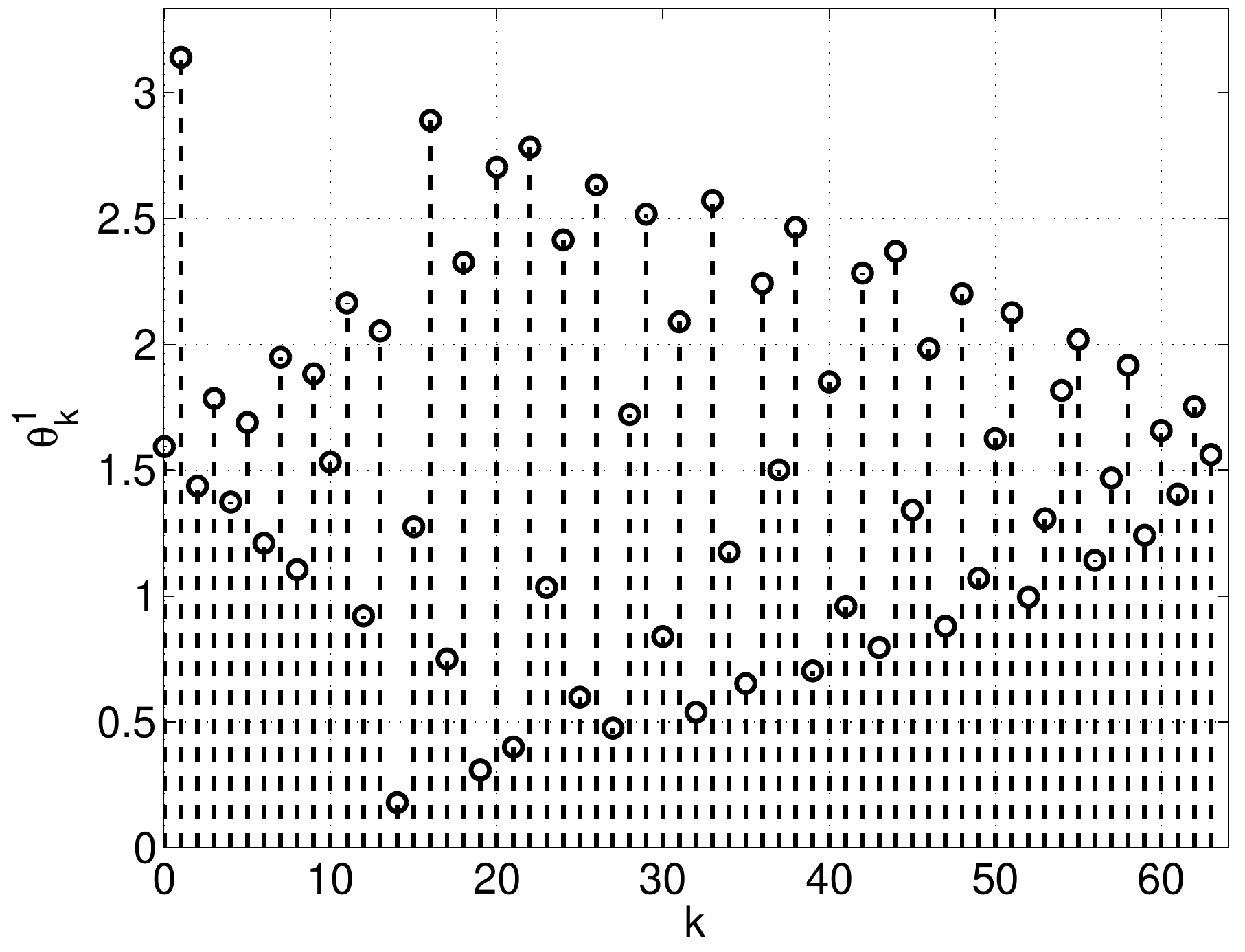}}\hfil
    \subfloat[]{
        \includegraphics[width=0.32\textwidth]{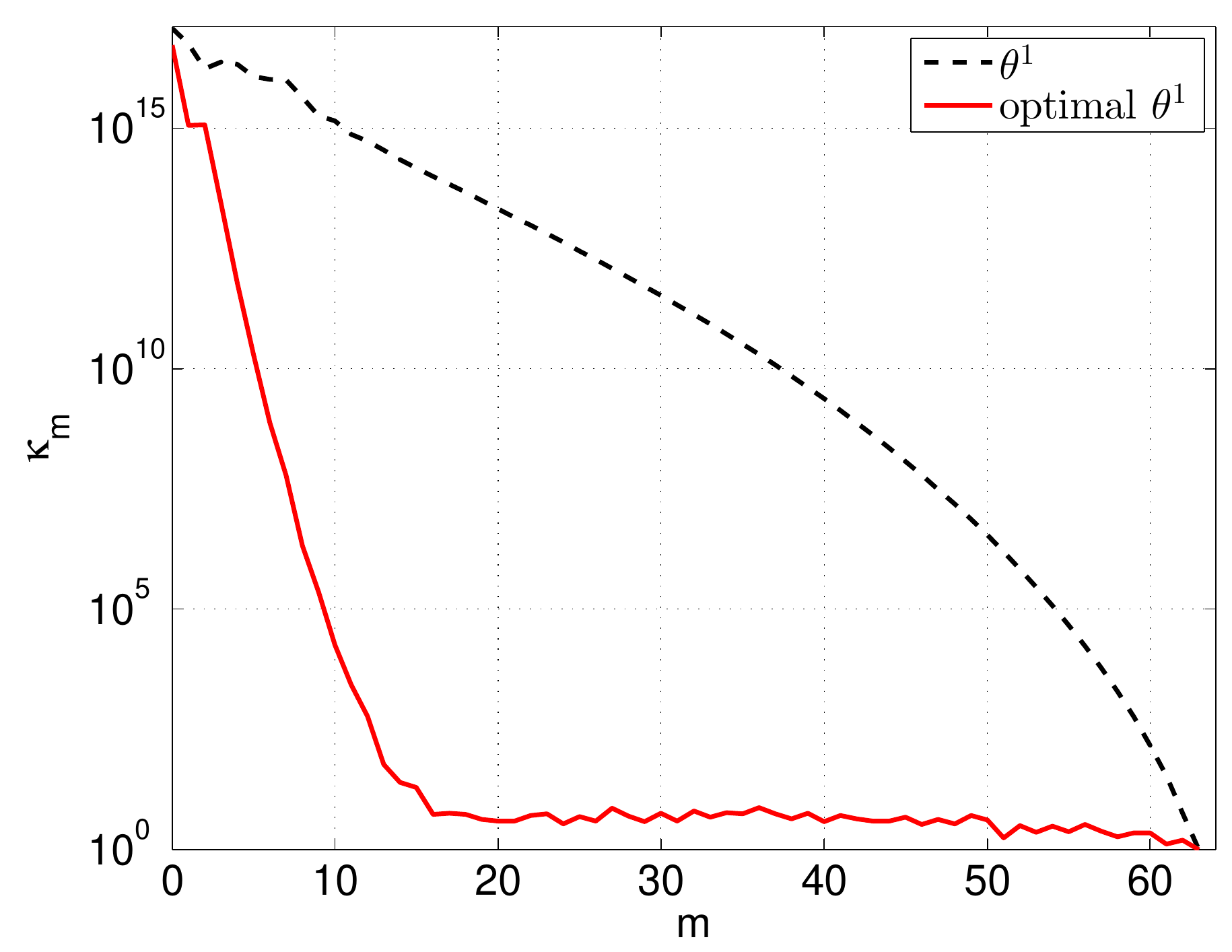}}\hfil
    \caption{(a) The sample positions $\bv{\theta^1}$ constructed from the set $\Theta^1$ of samples placed along co-latitude according to the measure $\sin\theta d\theta$ for the band-limit $L=64$,
      (b) the optimal sample positions $\bv{\theta^1}$ and (c) the condition number $\kappa_m$ of the matrix $\mathbf{P}^m$ for different
      values of $0 \leq m<L$, where the matrix $\mathbf{P}^m$ is constructed with the sample positions $\bv{\theta^1}$~(shown in (a)) or optimal sample positions $\bv{\theta^1}$~(shown in (b)) as indicated.
      }
    \label{fig:measure_sin}
\end{figure*}

For $L=64$, we show the sample positions $\bv{\theta^1}$ in \figref{fig:measure_sin}(a) and the condition number $\kappa_m$ of the matrix $\mathbf{P}^m$, constructed with the sample positions $\bv{\theta^1}$ for different values of $0 \leq m<L$ in \figref{fig:measure_sin}(c), where it can be observed that the placement of rings according to the measure $\sin\theta d\theta$ results in greater ill-conditioning of the $\mathbf{P}^m$ matrices as compared to the placement of rings according to the uniform measure $d\theta$. We also optimize the sample positions $\bv{\theta^1}$ by applying the proposed condition number minimization method. The optimal sample positions $\bv{\theta^1}$ are shown in \figref{fig:measure_sin}(b) and the condition number $\kappa_m$ of the matrix $\mathbf{P}^m$, obtained using the optimal sample positions $\bv\theta^1$ for different values of $0 \leq m<L$, is also plotted in \figref{fig:measure_sin}(c). Since the matrices $\mathbf{P}^m$ for the original $\bv\theta^1$ are highly ill-conditioned, the proposed condition number minimization method, that performs the re-ordering of the sample positions along co-latitude, does not find an ordering that significantly improves the ill-conditioning of the $\mathbf{P}^m$ matrices.

\begin{figure*}[t]
    \centering
    \hspace{-4mm}
    \subfloat[]{
        \includegraphics[width=0.32\textwidth]{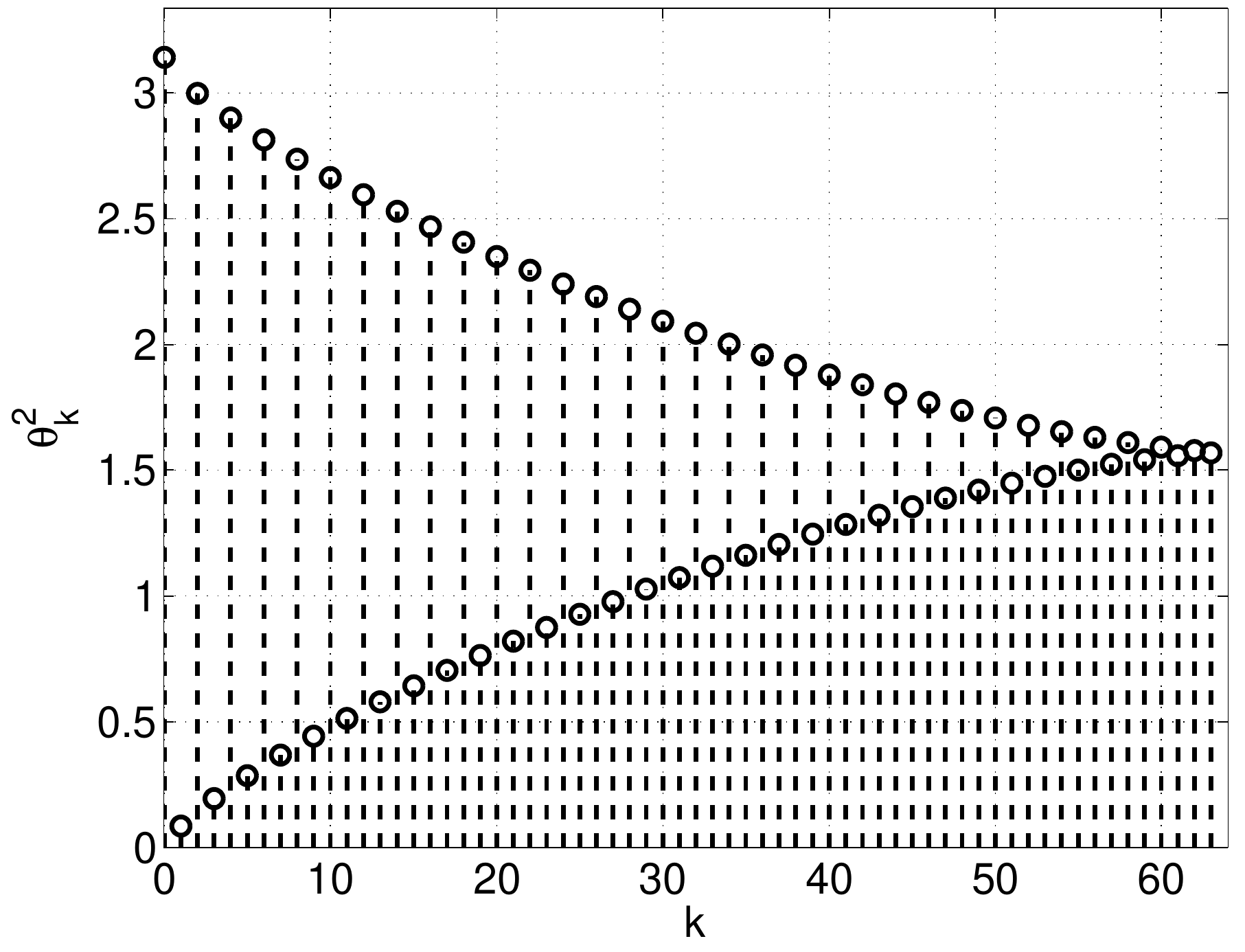}}\hfil
    \subfloat[]{
        \includegraphics[width=0.32\textwidth]{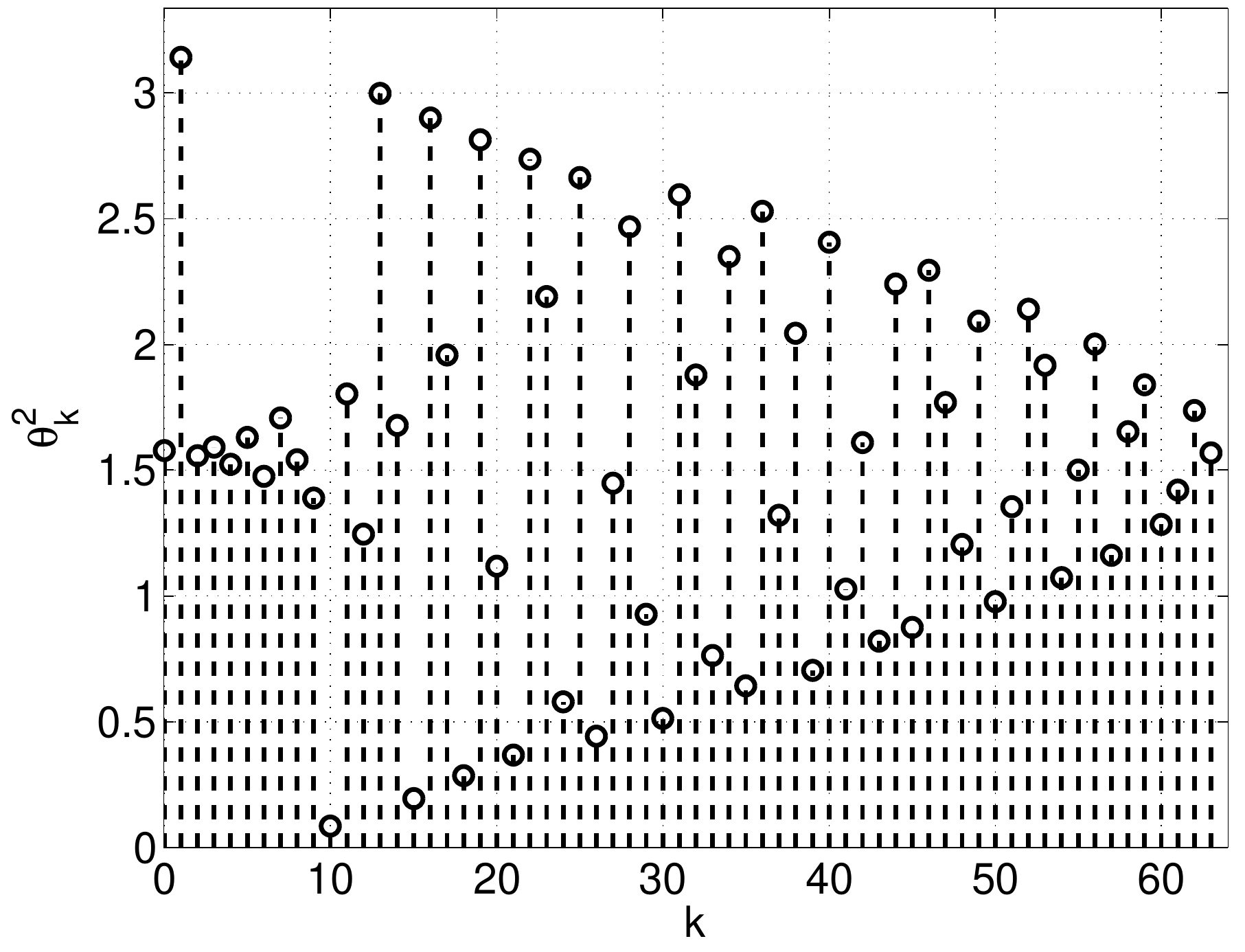}}\hfil
    \subfloat[]{
        \includegraphics[width=0.32\textwidth]{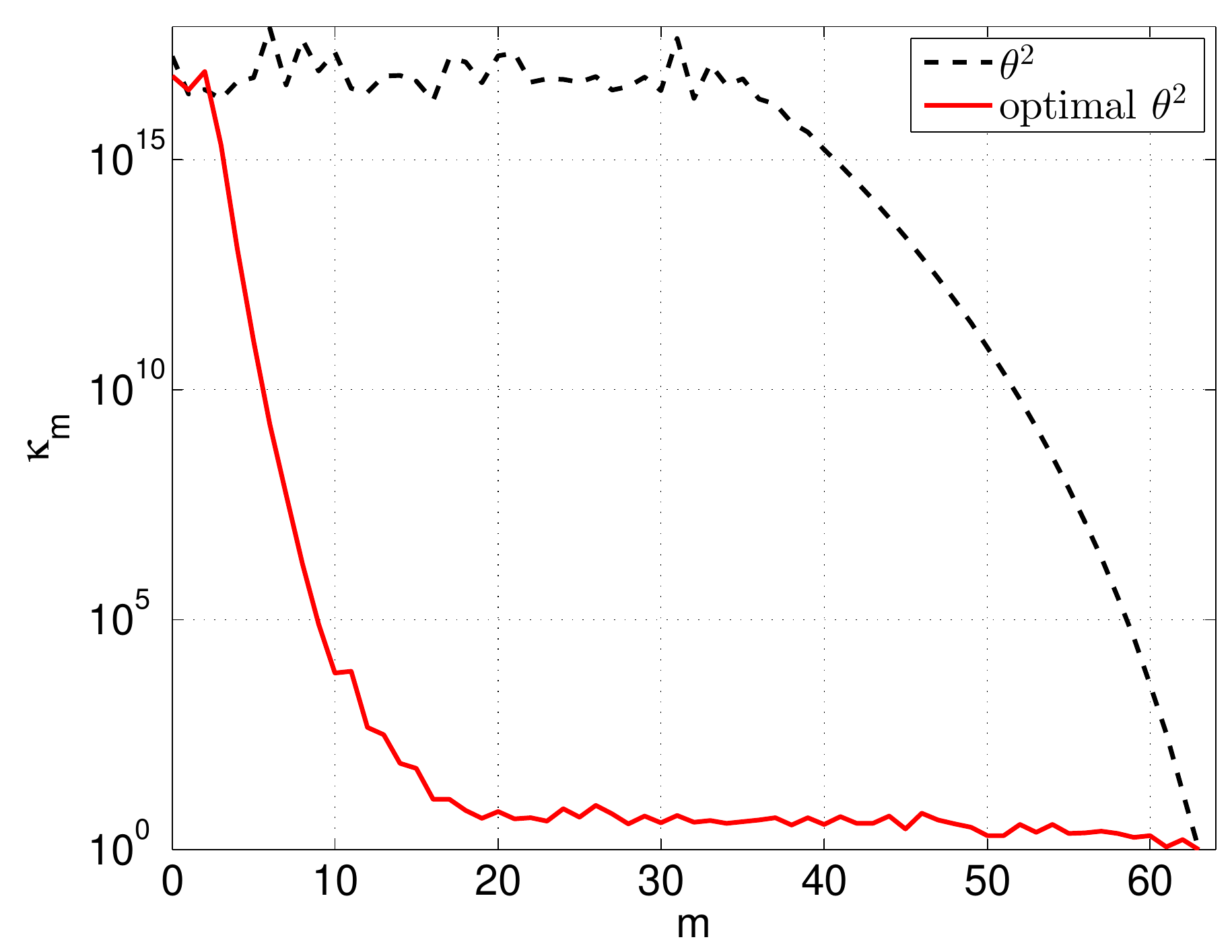}}\hfil
    \caption{(a) The sample positions $\bv{\theta^2}$ constructed from the set $\Theta^2$ of samples placed along co-latitude according to the measure $|\tan\theta|^{1/3}d\theta d\phi$ for the band-limit $L=64$,
      (b) the optimal sample positions $\bv{\theta^2}$ and (c) the condition number $\kappa_m$ of the matrix $\mathbf{P}^m$ for
      different values of $0 \leq m<L$, where the matrix $\mathbf{P}^m$ is constructed with the sample positions $\bv{\theta^2}$~(shown in (a)) or optimal sample positions $\bv{\theta^2}$~(shown in (b)) as indicated.
      }
    \label{fig:measure_tan}
\end{figure*}

We also carry out a similar analysis for the sample positions $\bv{\theta^2}$ shown in \figref{fig:measure_tan}(a). The optimal sample positions $\bv{\theta^2}$ obtained by applying the condition number minimization method is shown in \figref{fig:measure_tan}(b) and the condition number $\kappa_m$ of the matrix $\mathbf{P}^m$ using the sample positions $\bv{\theta^2}$ or optimal sample positions $\bv{\theta^2}$ is plotted in \figref{fig:measure_tan}(c) for different values of $0 \leq m<L$, which illustrates that the placement of rings according to the measure $|\tan\theta|^{1/3}d\theta$ also results in greater ill-conditioning of the $\mathbf{P}^m$ matrices as compared to the placement of rings according to the uniform measure $d\theta$. Thus, we conclude that the use of equiangular placement~(with uniform measure $d\theta$) of samples along co-latitude in the proposed sampling scheme performs well compared to the use of sampling methods which place the samples according to the measures $\sin\theta d\theta$ and $|\tan\theta|^{1/3}d\theta$. We note that the equiangular placement of samples is not the only choice to place samples along co-latitude and more sophisticated sampling methods can be developed. For example, the sampling scheme that also takes into account the location of samples in each ring along longitude can be designed. However, the development of such a design is beyond the scope of the current paper and is considered as a direction for future research.


\subsection{Computation of Spherical Harmonics}
The implementation of both forward and inverse transforms require the computation of scaled associated Legendre functions $\plms_\ell^m(\theta_k)= Y_\ell^m(\theta_k,0)$ for all degrees $\ell<L$ and orders $|m|\leq \ell$ and for all $\theta_k \in \bv{\theta}$. Different recursion relations can be used for the computation of associated Legendre functions for given $\theta_k$. For example, we can use the recursion proposed by Risbo~\cite{Risbo:1996} that computes $\plms_\ell^m(\theta_k)$ for all orders $|m|\leq\ell$ for given degree $\ell$ in each step of recursion, or alternatively, we can use the three-term recursion which grows with degree $\ell$ and recursively computes $\plms_\ell^m(\theta_k)$ for all $|m| \leq \ell <L-1$ for a given $m$.

We note that the forward transform iterates over different values of $m$ and uses $\mathbf{P}_m$~(which is composed of $\plms_\ell^m(\theta_k)$ of different $|m|\leq\ell<L$ and $\theta_k\in\bv{\theta}^{|m|}$) in each iterative step for the computation of spherical harmonic coefficients in a vector $\mathbf{f}_m$~(or $\mathbf{f}_{-m}$) as given in \eqref{Eq:gtof_inverse}. Furthermore, the computation of $G_m(\theta_k)$ in the implementation of the inverse transform also requires $\plms_\ell^m(\theta_k)$ for all $|m| \leq \ell <L$ for a given $m$. Therefore the three-term recursion is a natural choice to compute associated Legendre functions in our proposed transforms. For given $m$ and $\theta_k$, the three term recursion relation is given by
\begin{multline}
\label{Eq:three_term_recurrence}
    \sqrt{ \frac{ \big((\ell+1)^2 -m^2\big)  (2\ell-1) } {2\ell+1}}\, \plms_{\ell+1}^{m}(\theta_k) \\
    = \sqrt{(2\ell+3)(2\ell+1)}\, \cos\theta_k \,\plms_{\ell}^{m}(\theta_k) - \\
    \frac{\ell}{\ell+1} \sqrt{ \frac{ (\ell^2 -m^2) (2\ell+3) }{2\ell+1} }\, \plms_{\ell-1}^{m}(\theta_k),
\end{multline}
which grows with the degree $\ell$ for given $m$ with the following initial condition for $m\geq0$
\begin{align}\label{Eq:base_condition}
\plms_m^m(\theta_k)  &= (-1)^m\, \frac{\sqrt{(2m)!}}{2^m\,m!}\, \big(\sin\theta_k\big)^m
\end{align}
and symmetry relation which follows from (32) given in the appendix
\begin{align}
\plms_{\ell}^{-m}(\theta_k)  &= (-1)^m \plms_\ell^m(\theta_k).
\end{align}
The variant of recursion relation in \eqref{Eq:three_term_recurrence} has been adopted in the literature for the computation of associated Legendre polynomials~\cite{Healy:2003,Kostelec:2008}.
As demonstrated in~\cite{Reinecke:2013}, the three-term recurrence relation in \eqref{Eq:three_term_recurrence} is stable when the recurrence is carried out in the direction of increasing $\ell$, provided the initial condition in \eqref{Eq:base_condition} is computed accurately~(either by using higher than double precision arithmetic or by an adaptive rescaling).

\subsection{Extension to Spin Functions on Sphere}

By comparing the expansion of a spin function $\fs$ into spin spherical harmonics in \eqref{Eq:f_spin_expansion} with the expansion of the non-spin standard function $f$ in \eqref{Eq:f_expansion}, we note that the forward and inverse transforms developed for non-spin functions are also applicable to band-limited spin functions $\fs \in {}_s\lsphL{L}$ with the following associations
\begin{equation}
f \rightarrow \fs, \quad  {\shc{f}{\ell}{m}} \rightarrow {\shc{\fs}{\ell}{m}},\quad Y_\ell^m \rightarrow \ylms{\ell}{m}.
\end{equation}
The extension of forward and inverse transforms to the spin functions require the computation of spin weighted spherical harmonics $\ylms{\ell}{m}(\theta)\equiv \ylms{\ell}{m}(\theta,0)$, which can be carried out using the recurrence relation given in \appref{App:maths}, which is the generalized version of the recurrence relation in \eqref{Eq:three_term_recurrence}.  We do not further investigate the application of our transforms to spin function and limit our explorations for the non-spin standard functions in the rest of the paper.

\section{Analysis of Proposed Sampling Scheme and Transforms}\label{sec:analysis}

In this section we evaluate the proposed sampling scheme and associated SHTs using the following criteria: (1) the number of samples required to accurately represent a band-limited signal; (2) the computational complexity of the associated forward and inverse transforms; and (3) the accuracy of the forward and inverse transforms. We have carried out the implementation of the proposed transforms in double precision arithmetic. The code to compute the scaled associated Legendre function $\plms_\ell^m(\theta) \equiv Y_\ell^m(\theta,0)$ for given $\theta$, $\ell$ and $m$ using the recursion relation in \eqref{Eq:three_term_recurrence} is written in \clang\ in order to speed up the computation, and uses an adaptive rescaling so that double precision arithmetic is accurate.  The forward and inverse transforms, outlined as procedures in the previous section, are implemented in \matlab.

\subsection{Numerical Accuracy}
We analyse the numerical accuracy of our forward and inverse transforms that implement our proposed optimal sampling scheme on the sphere. The accuracy of the proposed transforms means that the inverse~(or forward) SHT of any band-limited signal followed by the forward~(or inverse) SHT yields the same band-limited signal, with error on the order of the numerical precision. In order to evaluate the numerical accuracy of the proposed SHTs, we carry out two numerical experiments, before further studying the error distribution in harmonic space.

\subsubsection{Experiment 1~(Spectral-Spatial-Spectral)}
\begin{figure}[t]
    \centering
    \vspace{-3mm}
    \hspace{-6mm}
    \includegraphics[scale=0.45]{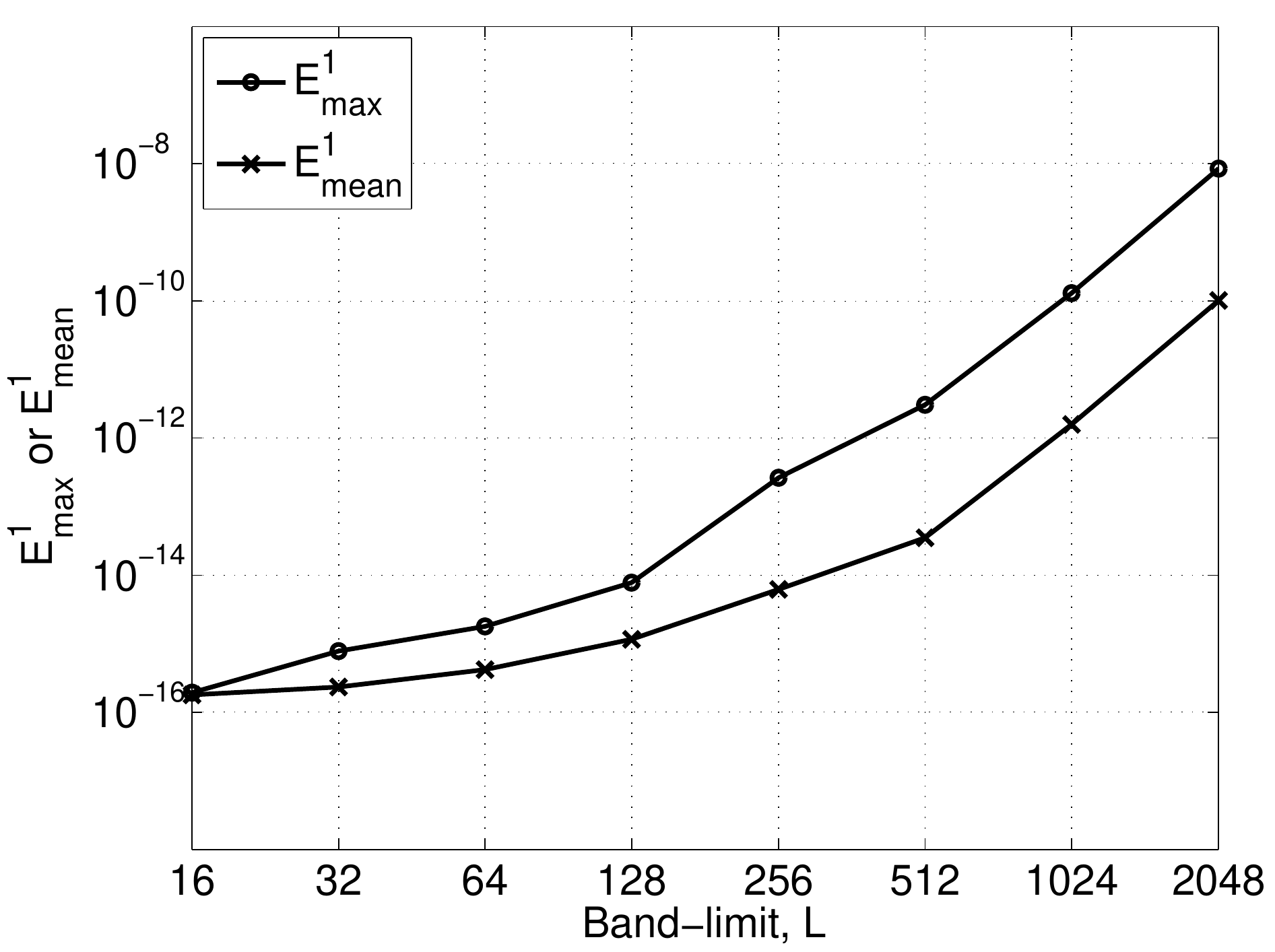}
    \caption{Experiment $1$ to test the numerical accuracy of the proposed transforms: for band-limits in the range $16\leq L\leq 2048$, we plot the maximum error and the mean error, respectively given in \eqref{Eq:exp1:errors:max} and \eqref{Eq:exp1:errors:mean}, between the test signal $f_{\rm t}$ and the reconstructed signal $f_{\rm r}$ in the spectral domain.}
    \label{fig:accuracy:one}
\end{figure}

In our first experiment, we generate the test signal spherical harmonic coefficients $\shc{f_{\rm t}}{\ell}{m}$ for $0<\ell<L, |m|\leq\ell$ with real and imaginary parts uniformly distributed in the interval $[-1,\,1]$. The inverse SHT is then used to synthesise the band-limited test signal $f_{\rm t}\in\lsphL{L} $ in the spatial domain over the $L^2$ samples of our sampling scheme, followed by the forward SHT to compute the reconstructed spherical harmonic coefficients, denoted by $\shc{f_{\rm r}}{\ell}{m}$. The experiment is repeated 10 times and the average values for the maximum error $E^1_\textrm{max}$ and the mean error $E^1_\textrm{mean}$, given by
\begin{align}\label{Eq:exp1:errors:max}
E^1_\textrm{max} &\dfn \max|\shc{f_{\rm t}}{\ell}{m} - \shc{f_{\rm r}}{\ell}{m} |, \\ E^1_\textrm{mean}  &\dfn \frac{1}{L^2} \sum_{\ell=0}^{L-1} \sum_{m=-\ell}^\ell |\shc{f_{\rm t}}{\ell}{m} - \shc{f_{\rm r}}{\ell}{m}| \label{Eq:exp1:errors:mean},
\end{align}
are recorded and plotted in \figref{fig:accuracy:one} for band-limits in the range $16\leq L\leq 2048$.

\begin{figure}[t]
    \centering
    \vspace{-3mm}
    \hspace{-6mm}
    \includegraphics[scale=0.45]{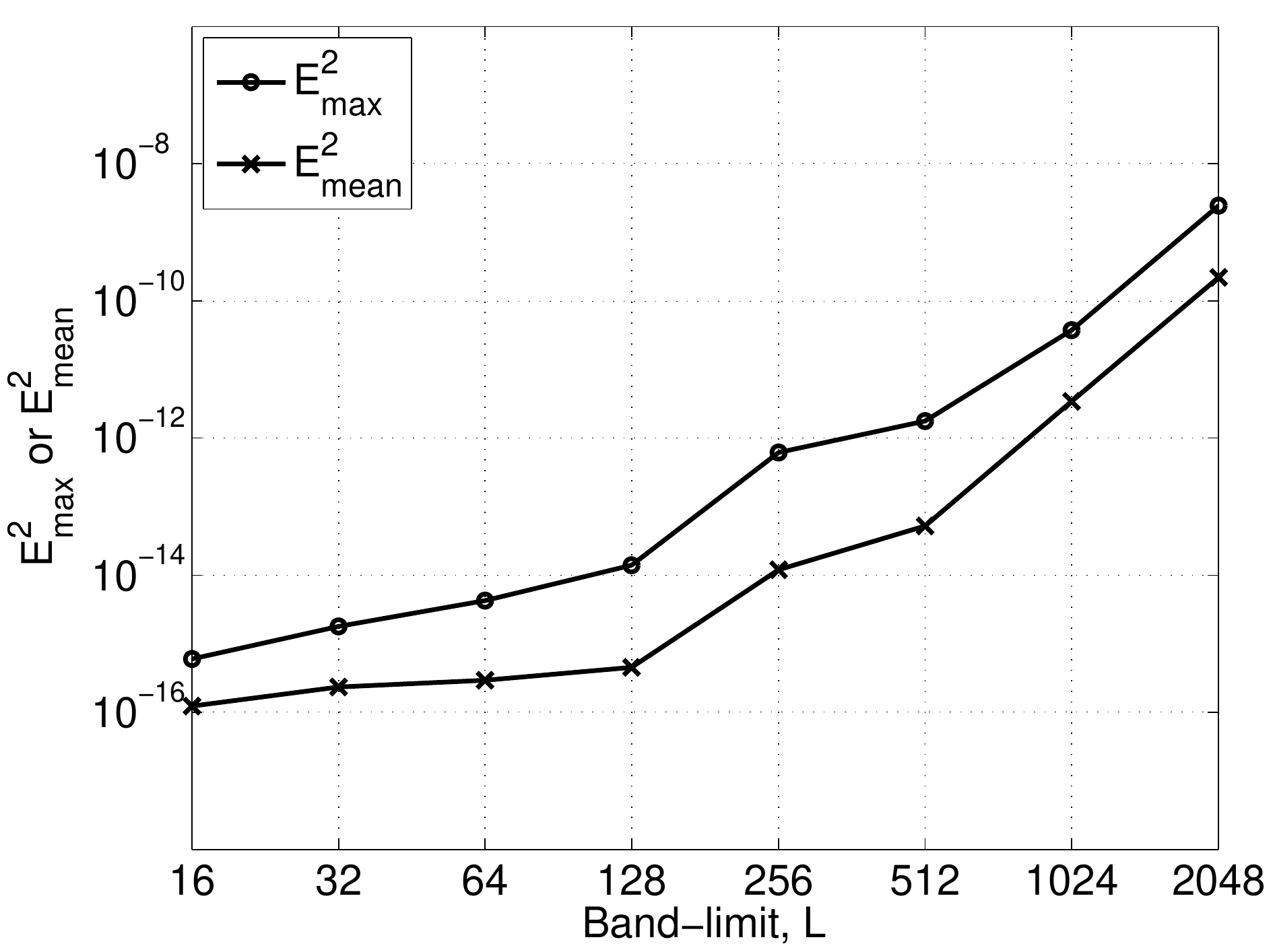}
    \caption{Experiment $2$ to test the numerical accuracy of the proposed transforms: for band-limits in the range $16\leq L\leq 2048$, we plot the maximum error and the mean error, respectively given in \eqref{Eq:exp2:errors:max} and \eqref{Eq:exp2:errors:mean}, between the test signal $f_{\rm t}$ and the reconstructed signal $f_{\rm r}$ in the spatial domain.}
    \label{fig:accuracy:two}
\end{figure}
\begin{figure}[t]
    \centering
    \vspace{-3mm}
    \hspace{-6mm}
    \includegraphics[scale=0.62]{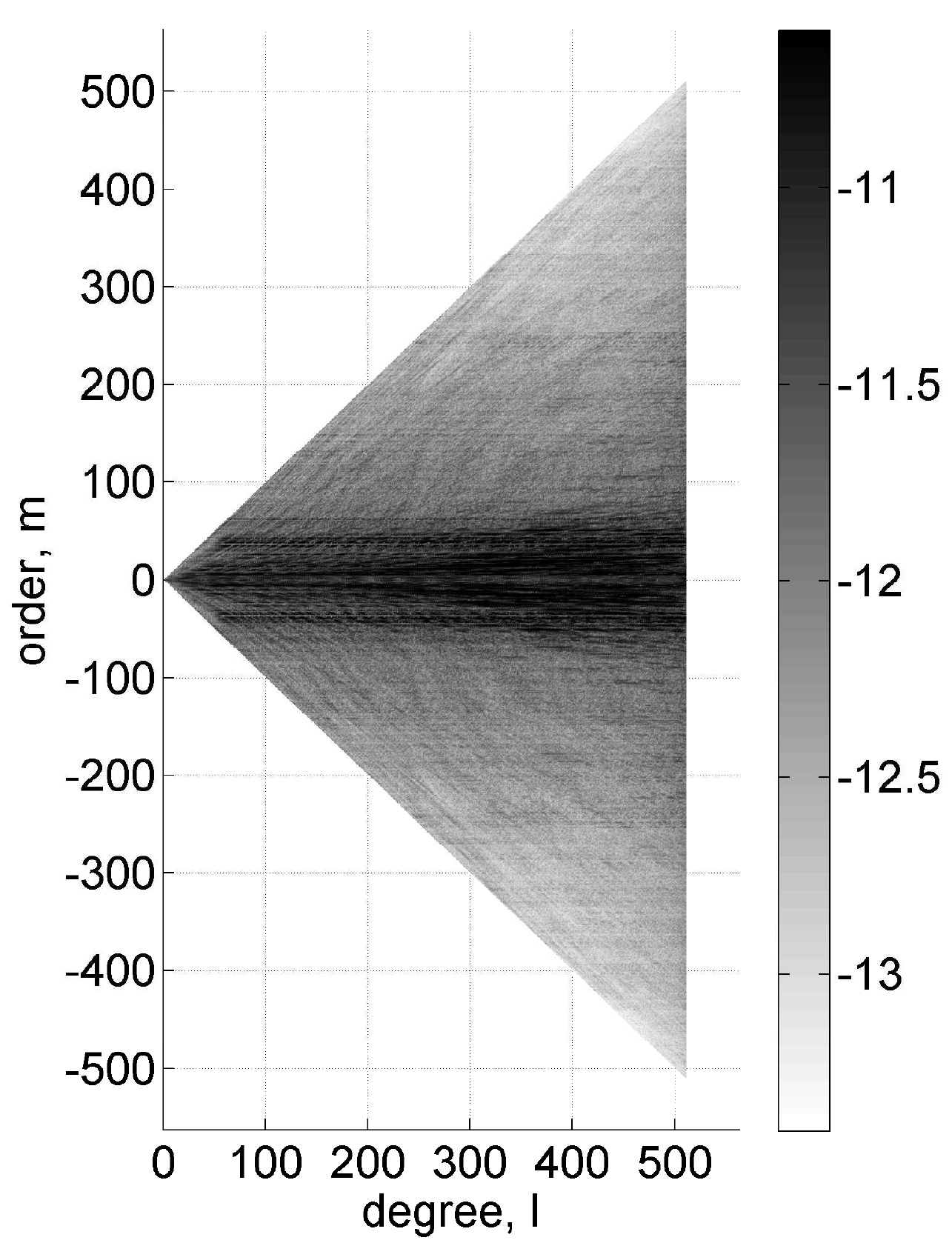}
    \caption{For experiment $1$,  the error $E_\ell^m = |\shc{f_{\rm t}}{\ell}{m} - \shc{f_{\rm r}}{\ell}{m}|$ between the spherical harmonic coefficients of the test signal and coefficients of the reconstructed signal. $E_\ell^m$ is plotted in base-10 logarithmic scale as a surface plot for all degrees $\ell<L$ and orders $|m|\leq \ell$, where the band-limit is $L=512$. Note that the error increases as order $|m|$ decreases from $L-1$ to $0$.}
    \vspace{-2mm}
    \label{fig:accuracy:three}
\end{figure}

\subsubsection{Experiment 2~(Spatial-Spectral-Spatial)}
In the second experiment to test the numerical accuracy of the proposed transforms, we randomly generate the complex valued band-limited test signal $f_{\rm t}\in\lsphL{L}$ with real and imaginary parts uniformly distributed in the interval $[-1,\,1]$ over the $L^2$ samples of our sampling scheme. The forward SHT, followed by inverse SHT is applied on the test signal to obtain the reconstructed signal $f_{\rm r}\in\lsphL{L}$. We repeat the experiment 10 times and the obtain the average values for the maximum error $E^2_\textrm{max}$ and the mean error $E^2_\textrm{mean}$ between the test signal $f_{\rm t}$ and reconstructed signal $f_{\rm r}$, defined as
\begin{align}\label{Eq:exp2:errors:max}
E^2_\textrm{max} &\dfn \max|f_{\rm t}(\theta,\phi) - f_{\rm r}(\theta,\phi)|, \\ E^2_\textrm{mean} &\dfn \frac{1}{L^2} \sum_{(\theta,\phi)} |f_{\rm t}(\theta,\phi) - f_{\rm r}(\theta,\phi)| \label{Eq:exp2:errors:mean},
\end{align}
where the sum is over all $(\theta,\phi)$ points in the proposed sampling scheme. Both $E^2_\textrm{max}$ and $E^2_\textrm{max}$ are plotted in \figref{fig:accuracy:two} for the band-limit in the range $16\leq L\leq 2048$.

\subsubsection{Further Analysis}
It can be observed that both the maximum error $E_\textrm{max}$ and the mean error $E_\textrm{mean}$ grows quadratically with the band-limit $L$, which is due to the computational flow of the proposed transform.  The proposed transform sequentially computes the spherical harmonic coefficients $\shc{f}{\ell}{m}$, first for order $|m|=L-1$, proceeding to order $|m|=0$. The computation of the spherical harmonic coefficients of order $m$ requires knowledge of all coefficients with order greater than $|m|$ because the forward transform eliminates the effect of all coefficients greater than $m$ from the sample positions in the rings placed at $\bv\theta \backslash \bv\theta^m $ to avoid aliasing, before taking FFTs along these rings. Any error introduced in the computation of the spherical harmonic coefficients of order $|m|$ propagates in the computation of coefficients of order less than $|m|$. In order to further elaborate, we plot the error $E_\ell^m = |\shc{f_{\rm t}}{\ell}{m} - \shc{f_{\rm r}}{\ell}{m}|$ for $L=512$ and averaged over 10 realization of experiment $1$ in \figref{fig:accuracy:three}, where it can be observed that error is comparatively smaller for higher order coefficients and increases as order $m$ decreases from $L-1$ to $0$.

\subsection{Why is Spatial Dimensionality Important?}
The fundamental property of any sampling scheme is the number of samples required to accurately represent a band-limited signal. The existing sampling schemes in the literature~\cite{Skukowsky:1986,Doroshkevich:2005,Driscoll:1994,Healy:2003,McEwen:2011} that support an accurate SHT do not attain the optimal spatial dimensionality $\nps=L^2$ on the number of samples, as also highlighted earlier. In comparison, our proposed sampling scheme and associated transforms require $\nps = L^2$ samples, which is the optimal spatial dimensionality attainable by any sampling scheme since the band-limited signal belongs to the $L^2$ dimensional subspace $\lsphL{L}$.

Now, we briefly discuss the significance of achieving optimal spatial dimensionality. In addition to the practical considerations~\cite{Zhang:2012,Tuch:2004,McEwen:2013} that desire fewer samples for the representation of band-limited signals,  we highlight that experiment 2, composed of forward transform of a signal \emph{randomly} generated over $\nps = L^2$ samples followed by the inverse transform, yields the original signal in the spatial domain.  This is not the case for existing sampling schemes since a random signal over $nL^2$ samples in the spatial domain, where typically $n\sim2$ \cite{McEwen:2011} or $n\sim4$ \cite{Driscoll:1994}, may not belong to the $L^2$ dimensional subspace of band-limited signals with band-limit $L$.

\subsection{Computational Complexity Analysis}

First, we analyse the computational complexity of the proposed forward SHT. Following the forward SHT procedure, the complexity to compute $\mathbf{g}_m$ for each $m$, which only requires one $2m+1$ point FFT, is $O(L\log L)$. Using the recursive relation in \eqref{Eq:three_term_recurrence}, $\plms_\ell^m(\theta_k)$ for all $|m| \leq \ell <L$ and for all $\theta_k\in\bv{\theta}$ and for each $m$ and can be computed in $O(L^2)$ time.

Since the computation of $\mathbf{f}_m$ requires solving the system in \eqref{Eq:gtof_inverse}, that can be carried out naively using the least squares approach with complexity $O(L^3)$. However, the system in \eqref{Eq:gtof_inverse} can be solved more efficiently in practice by employing fast algorithms. For example, the system of size $L$ can be solved in $O(L^{2.37})$, instead of $O(L^3)$, using the algorithm of \cite{Coppersmith:1987}. Once $\mathbf{f}_m$ is computed, the effect of higher order spherical harmonics is removed from the signal, which can be carried out in two steps: 1) evaluation of $\tilf_m(\theta,\phi)$ given in \eqref{Eq:signal_subset}, which can be done asymptotically in $O(L^2)$ for each $m$; and (2) updating the signal as $f(\theta,\phi) \leftarrow f(\theta,\phi)-\tilf_m(\theta,\phi)$ for all $\theta_k \in \bv\theta \backslash \bv\theta^m $ and all associated sampling points along $\phi$, which is again performed with complexity $O(L^2)$. Since these operations need to be repeated for each $m$, and $m$ is of the order $L$, the complexities mentioned above are scaled by $L$ and therefore the overall asymptotic complexity of the forward transform scales naively as $O(L^4)$. The dominant factor $O(L^4)$ is due to the inverting of system in \eqref{Eq:gtof_inverse} only, which can be more efficiently implemented in practice and we demonstrate later in this section that the complexity of both the inversion of the system and the forward transform scales close to $O(L^3)$ in practice. If the matrices $\mathbf{P}_m$ and the inverse matrices $\mathbf{P}_m^{-1}$~(or pseudo-inverse) for all $m<L$ are pre-computed, the theoretical complexity reduces to $O(L^3)$. However, the pre-computation requires the storage of the order $O(L^3)$ and quickly becomes infeasible for higher band-limits~\cite{Huffenberger:2010,McEwen:2011}.
We note that the complexity to compute the scaled associated Legendre function $\plms_\ell^m(\theta_k)$ for all $|m| \leq \ell <L$ and for all $\theta_k\in\bv{\theta}$ is $O(L^3)$, which is not greater than the computational complexity of the forward transforms and therefore the on-the-fly computation of spherical harmonics does not alter the overall complexity of the proposed forward or inverse SHT.

\begin{figure}[t]
    \centering
    \vspace{-3.5mm}
    \hspace{-6mm}
    \includegraphics[scale=0.45]{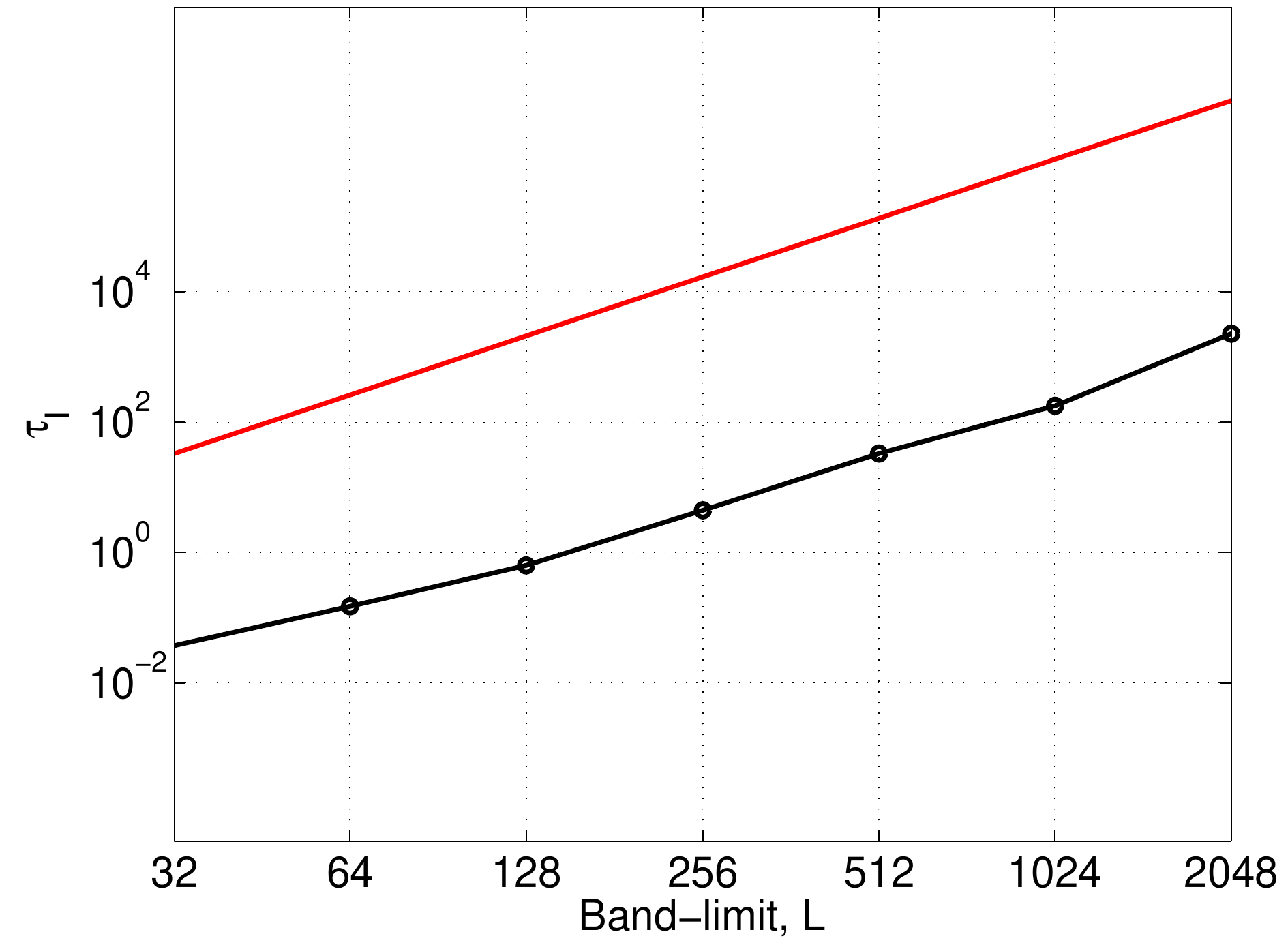}
    \caption{The computation time $\tau_I$~(in seconds) taken by the proposed inverse spherical harmonic transform to compute the complex signal of different band-limits in the range $32 \leq L \leq 2048$ from its spherical harmonic coefficients. Note that $\tau_{I}$ scales as $O(L^3)$ as indicated by red line~(without markers).}
    \label{fig:timing:inv}
\end{figure}
%

Following the inverse SHT procedure, we first compute scaled associated Legendre functions  $\plms_\ell^m(\theta_k)$ for each $m$ and for all $\ell$ and all $\theta_k$ in $O(L^2)$, which is then used in \eqref{Eq:gm_integral} to evaluate $G_m(\theta_k)$. Therefore the complexity to compute $G_m(\theta_k)$ for all $\theta$ and each $m$ is $O(L^2)$ and for all $m$ is $O(L^3)$. Once $G_m(\theta_k)$ is known, the summation over $m$ can be evaluated to compute the signal $f(\theta,\phi)$ at all spatial samples in $O(L^3)$. Thus, the inverse SHT has the overall complexity of $O(L^3)$, which is similar to the complexity of the transforms that exist in literature for different sampling schemes on the sphere.

\begin{figure}[t]
    \centering
    \vspace{-3.5mm}
    \hspace{-6mm}
    \includegraphics[scale=0.45]{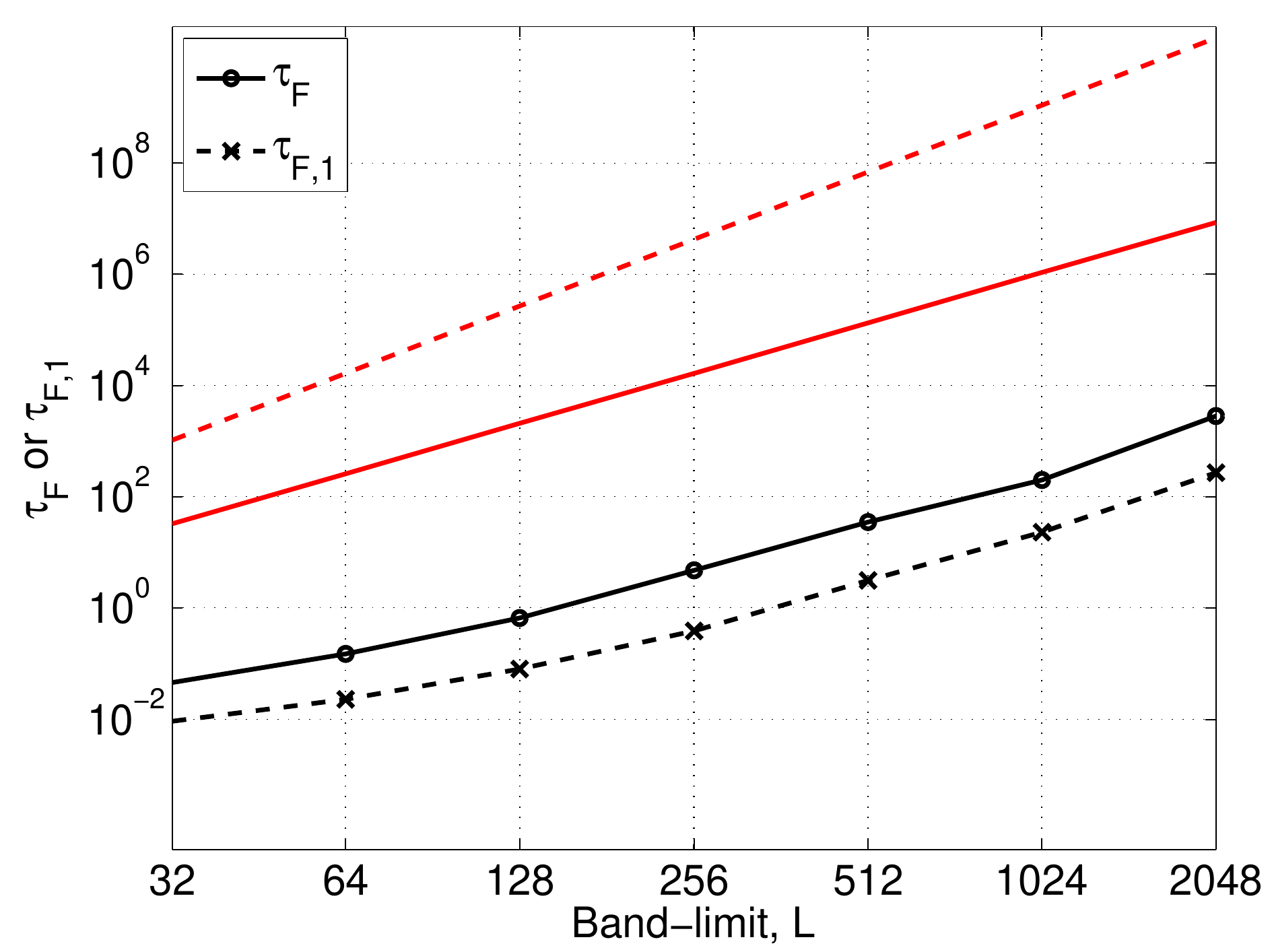}
    \caption{The computation times $\tau_{F}$ and $\tau_{F,1}$ in seconds. The time $\tau_{F}$ is taken by the proposed forward spherical harmonic transform to compute the spherical harmonic coefficients of the complex signal of different band-limits in the range $32 \leq L \leq 2048$. The computation time $\tau_{F,1}$ taken by step $5$ of the forward spherical harmonic transform procedure. Note that both $\tau_{F}$ and $\tau_{F,1}$ scale close to $O(L^3)$ instead of $O(L^4)$ due to the use of efficient techniques for inverting the matrix system in practice. The $O(L^3)$ and $O(L^4)$ scaling is shown by solid and dashed red lines~(without markers) respectively.  }
    \label{fig:timing:fwd}
\end{figure}

We measure the computation times, denoted by $\tau_I$ and $\tau_F$ to carry out the proposed inverse and forward transforms, respectively, in experiment 1 detailed in the previous subsection for band-limits $32 \leq L\leq 2048$. We also record the computation time, denoted by $\tau_{F,1}$, to perform only the step $5$ of the forward SHT procedure, which involves solving the matrix system of the form given in \eqref{Eq:gtof_inverse} and is the only step that makes the theoretical computational complexity of $O(L^4)$ for the forward SHT. An experiment is performed using \texttt{MATLAB} running on a machine equipped with 2.6 GHz Intel Core i7 processor and 64 GB of RAM and the computation times are averaged over 10 test signals. The computation time $\tau_I$, plotted in \figref{fig:timing:inv}, evolves as $O(L^3)$ as dictated by the red solid line~(without markers). The computation times $\tau_{F}$ and $\tau_{F,1}$ are plotted in \figref{fig:timing:fwd} where it can be noted that the both $\tau_{F}$ and $\tau_{F,1}$ scale closer to $O(L^3)$ instead of $O(L^4)$ in practice for band-limits up to $L =  2048$, which is due to the use of computationally optimized routines in LAPACK used by \texttt{MATLAB}.

A least squares system of size $L^2$ can also be constructed based on the proposed sampling scheme. Since the least squares separation of variables approach requires $2L^2$ number of samples, that is, $2L$ samples in each of the $L$ iso-latitude rings to avoid aliasing errors, it cannot be used for the proposed sampling scheme. Due to this fact, the complexity to solve a least squares system for the proposed sampling scheme scales with $O(L^6)$, which makes the least squares approach computationally infeasible even for smaller band-limits. Furthermore, reconstruction error using least squares with $L^2$ samples is poor even for low band-limits, e.g. reconstruction error for band-limit $L=64$ is $10^{-6}$ and the error grows with the band-limit.
In the proposed method, it is the removal of the contribution of coefficients of order greater than $m$ from the sample positions in the rings placed at $\bv\theta \backslash \bv\theta^m $ before taking FFT along these rings, which enables the elimination of aliasing errors and the efficient and accurate computation of the SHT.

\vspace{-2mm}
\subsection{Potential Applications}
We discuss three potential applications of our proposed sampling scheme and associated SHTs in the fields of acoustics, medical imaging and compressive sampling.

In acoustics, the head-related transfer function (HRTF), which serves as a quantitative measure of the response of human body anatomical features to sound waves, is required in the reconstruction of real life auditory scenes and is determined by setting up an experiment to take measurements over the sphere~\cite{Zhang:2010,Zhang:2012}. The HRTF is a band-limited function on the sphere, where the band-limit $L$ varies directly with the audio frequency and the band-limit corresponding to the maximum frequency of 20kHz is $L\sim47$~\cite{Zhang:2012}. Since the measurements over the sphere involves the rotation of a sound source or listener or both, a sampling scheme which requires fewer samples implies that the band-limited HRTF function can be measured exactly for lower cost. Since our sampling scheme achieves the optimal limit on the number of samples, it can potentially be adopted for taking measurements and accurate HRTF representation. We note that the proposed sampling scheme also has flexibility in terms of sample positions along longitude as the sample positions can be flexibly rotated~(or placed) along the ring. Furthermore, the placement of samples along co-latitude given in \eqref{Eq:sampling_latitude} can be chosen because the maximum band-limit in acoustics is $L\sim47$, for which the maximum condition number of the matrix $\mathbf{P}_m$ with the consideration of $\bv{\theta}$ in \eqref{Eq:sampling_latitude} is only of the order $10^2$, resulting in an accurate computation of the SHT.

The reduction in number of samples required to represent band-limited function is of great importance in applications, where the cost of acquiring a single sample is large. For example, diffusion magnetic resonance imaging (MRI) in medical imaging is one such application, where the cost is measured in terms of sample acquisition time. The acquisition strategies consider sampling on multiple spherical shells for each voxel of the brain and are time consuming since millions of voxels are generally considered. The total number of samples, and thus total acquisition time, can be reduced by a factor of at least two when replacing the existing sampling methods with our proposed sampling scheme.  Such an enhancement in acquisition time is of considerable importance in order to make diffusion MRI accessible for clinical use. Furthermore, since the band-limit for each spherical shell is considered to be of the order $L\sim10$, the placement of samples along co-latitude given in \eqref{Eq:sampling_latitude} can be used directly for the exact representation of data on each spherical shell.

In compressive sampling~\cite{Candes:2006,Donoho:2006}, the ratio of the number of required measurements to the spatial dimensionality of the signal scales approximately linearly with its sparsity.  Since our proposed sampling scheme achieves the optimal spatial dimensionality, compared to the other schemes, it will increase the performance of compressed sensing reconstruction on the sphere when recovering signals directly on the sphere. Furthermore, for sparsity priors defined in the spatial domain, such as signals sparse in the magnitude of their gradient, sparsity is also directly related to the sampling of the signal \cite{McEwen:2013}, where our optimal sampling scheme is likely to provide further enhancement.  The use of our sampling scheme in compressed sampling reconstruction on the sphere is likely to have impact on a variety of problems including more efficient acquisition, denoising, extrapolation and deconvolution on the sphere.

\section{Conclusions}\label{sec:conclusions}

For the accurate representation of a signal on the sphere
band-limited at $L$ with $L^2$ degrees of freedom in the spectral
domain, the existing sampling schemes, which support accurate
computation of spherical harmonic transform, require $2L^2$ or
$4L^2$ samples.
We have proposed a new sampling scheme on the sphere which only requires $L^2$ samples to represent a band-limited signal. Thus, the proposed sampling scheme matches the spectral dimensionality of the signal. For the proposed sampling scheme, we have also developed forward and inverse spherical harmonic transforms, which allow the computation of transforms with sufficient accuracy and manageable computational complexity and do not require any precomputation associated with SHT once the sample positions are determined. We have conducted numerical experiments to show the stability, accuracy and computational complexity of the proposed transforms up to $L=2048$.

Our optimal dimensionally sampling scheme and associated spherical
harmonic transforms have great potential for use in practical
applications found in acoustics, cosmology, geophysics and beyond.
For example, the reduction in the number of samples in the proposed
scheme may be exploited to reduce the cost of acquisition
significantly in diffusion MRI. Since the choice of sampling along
latitude is adaptive in the proposed scheme, this additional
flexibility can be exploited in practical problems. For example, in
applications where measurements are only available in spatially
limited regions, the proposed sampling scheme can be adapted to
these regions.

\begin{appendices}

\section{Mathematical Background}
\label{App:maths}

\subsection{Spherical Harmonics}

The spherical harmonic function, $Y_{\ell}^m(\theta, \phi)$, for degree
${\ell} \geq 0$ and order $ |m| \leq {\ell}$ is defined as~\cite{Kennedy-book:2013,Sakurai:1994}
\begin{equation}
\label{Eq:Sph_harmonics}
    Y_{\ell}^m(\theta, \phi) =
    N_{\ell}^{m}P_{\ell}^{m}(\cos\theta)\,e^{im\phi},
\end{equation}
with $N_{\ell}^{m} \dfn \sqrt{\frac{2{\ell}+1}{4\pi}\frac{({\ell}-m)!}{({\ell}+m)!}}$ is the normalization factor
such that $\innerp[\big]{Y_{\ell}^m}{ Y_{p}^{q}}=\delta_{\ell, p} \delta_{m,q}$, where $\delta_{m,q}$ is the Kronecker delta function: $\delta_{m,q} = 1$ for $m=q$ and is zero otherwise. $P_{\ell}^{m}(x)$ is the associated Legendre function defined for degree $\ell$ and order $0 \leq m \leq \ell$ as
\begin{align*}
    P_{\ell}^{m}(x)
        &= \frac{(-1)^m}{2^{\ell} {\ell}!} (1-x^2)^{m/2} \frac{d^{{\ell}+m}}{dx^{{\ell}+m}}
            (x^2-1)^{\ell} \\
    P_{\ell}^{-m}(x)
        &= {(-1)^m} \frac{({\ell}-m)!}{({\ell}+m)!} P_{\ell}^m(x),
\end{align*}
for $|x|\leq 1$. We also note the following relation between $Y_\ell^m$ and $Y_\ell^{-m}$
\begin{equation}
\label{eq:yellm:conj}
    \conj{Y_\ell^m(\theta,\phi)} = (-1)^mY_\ell^{-m}(\theta,\phi).
\end{equation}

\subsection{Spin Spherical Harmonics}

The spin spherical harmonic functions~(or  spin spherical harmonics for short), denoted by $\ylms{\ell}{m}$ are defined for degree $\ell$, order $m$ and $|s|\leq\ell$ as
\begin{align}
\ylms{\ell}{m}(\theta,\phi) \dfn (-1)^s \sqrt{\frac{2\ell+1}{4\pi}} \, e^{im\phi} d_{\ell}^{m,-s}(\theta),
\end{align}
where $d_{\ell}^{m,m'}(\theta)$ denotes the Wigner-$d$ function~\cite{Sakurai:1994,Kennedy-book:2013}. The spin spherical harmonics $\ylms{\ell}{m}(\theta)$ can be computed by directly evaluating the Wigner-$d$ function. However, in practice, $\ylms{\ell}{m}(\theta)$ is computed using the recurrence relation for Wigner-$d$ functions~\cite{Kostelec:2008} as
\begin{align}\label{Eq:three_term_recurrence_Wigner}
\frac{1}{\ell+1} &\sqrt{  \frac   { \big((\ell+1)^2 -m^2\big) \big((\ell+1)^2 -s^2\big)(2\ell-1)    }     {2\ell+1}   }\, \ylms{\ell+1}{m}(\theta)s \nonumber \\
&=
 \bigg(\cos\theta + \frac{ms}{\ell(\ell+1)}\bigg)\sqrt{(2\ell+3)(2\ell+1)}\, \ylms{\ell}{m}
\nonumber \\
&-
\frac{1}{\ell+1} \sqrt{  \frac   { (\ell^2 -m^2) (\ell^2 -s^2)(2\ell+3)    }     {2\ell+1}   }\, \ylms{\ell-1}{m}
\end{align}
which grows with the degree $\ell$ for given $m$ and $s$ with following initial condition for $|m|<L$, $|s|<L$ and $s\leq m$
\begin{align}
\ylms{m}{m}  = (-1)^m \sqrt{\frac{(2m)!}{(m+s)!(m-s)!}} \bigg(\!\cos\frac{\theta}{2}\bigg)^{m+s} \!\! \bigg(\!\sin\frac{\theta}{2}\bigg)^{m-s}\!\!.
\end{align}

\end{appendices}



\begin{IEEEbiography}[{\includegraphics[width=1in,height=1.25in,clip,keepaspectratio]{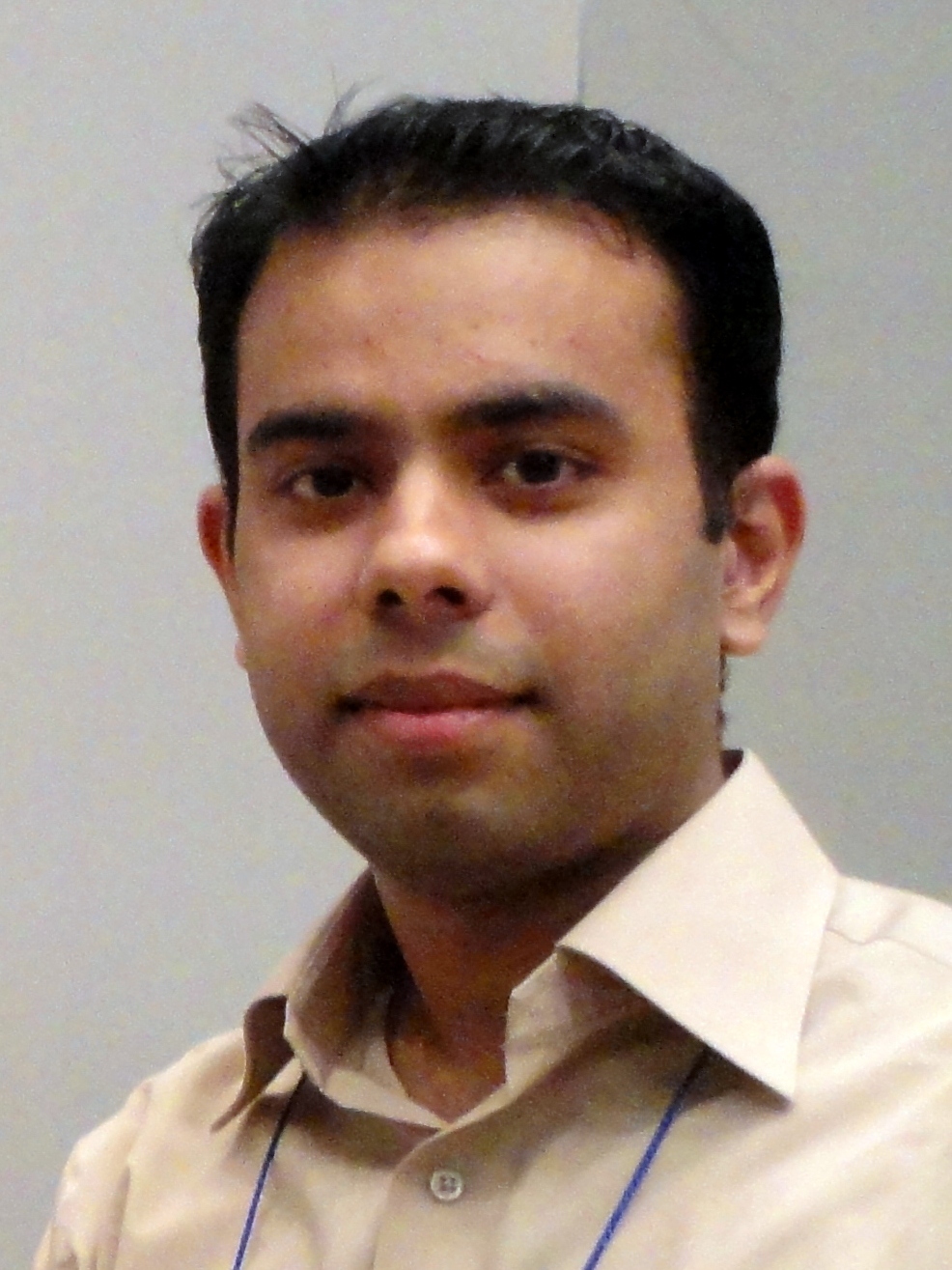}}]{Zubair Khalid}
(S'10-M'13) received his B.Sc. (First-class Hons.) degree in
Electrical Engineering from the University of Engineering $\&$
Technology (UET), Lahore, Pakistan in 2008. He received the Ph.D. degree in Engineering from the Australian National University of Canberra, Australia in Aug. 2013.

He is currently working as a Research Fellow in the Research School of Engineering, Australian National University~(ANU), Canberra, Australia. Zubair was
awarded University Gold Medal and Industry Gold Medals from Siemens
and Nespak for his overall outstanding performance in Electrical
Engineering during the his undergraduate studies. He was a recipient
of an Endeavour International Postgraduate Award for
his Ph.D. studies. His research interests are in
the area of signal processing and wireless communications, including the development of novel signal processing techniques for
signals on the sphere and the application of stochastic geometry in wireless ad-hoc networks .
\end{IEEEbiography}

\begin{IEEEbiography}[{\includegraphics[width=1in,height=1.25in,clip,keepaspectratio]{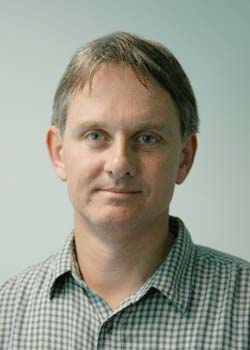}}]{Rodney A. Kennedy}
(S'86-M'88-SM'01-F'05) received the B.E. degree from the University
of New South Wales, Sydney, Australia, the M.E. degree from the
University of Newcastle, and the Ph.D. degree from the Australian
National University, Canberra.

He is currently a Professor in the Research School of Engineering,
Australian National University. He is a Fellow of the IEEE. His
research interests include digital signal processing, digital and
wireless communications, and acoustical signal processing.
\end{IEEEbiography}

\begin{IEEEbiography}[{\includegraphics[width=1in,height=1.25in,clip,keepaspectratio]
 {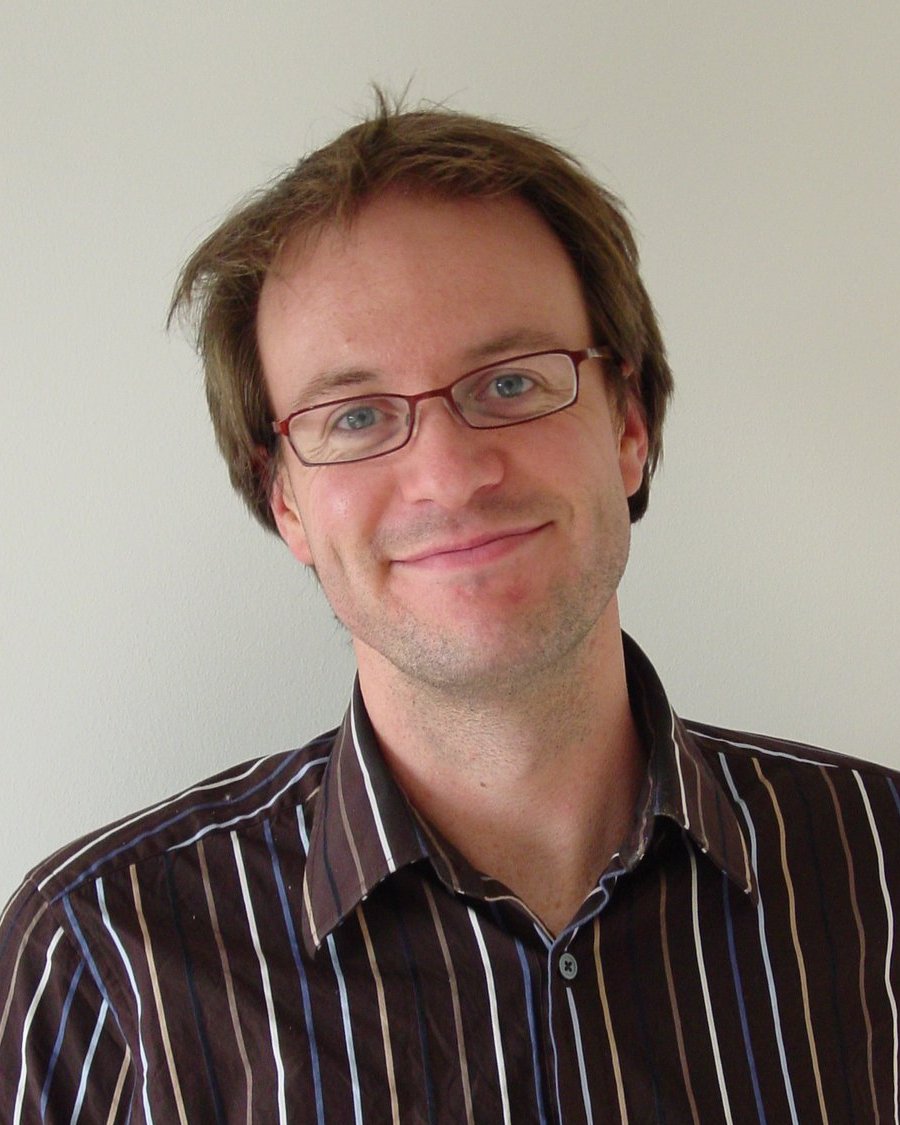}}]{Jason D. McEwen}
 received a B.E.\ (Hons) degree in Electrical and Computer Engineering from the University of Canterbury, New Zealand, in 2002 and a Ph.D.\ degree in Astrophysics from the University of Cambridge in 2007.

 He held a Research Fellowship at Clare College, Cambridge, from 2007 to 2008, worked as a Quantitative Analyst from 2008 to 2010, and held a position as a Scientist at Ecole Polytechnique F{\'e}d{\'e}rale de Lausanne (EPFL), Switzerland, from 2010 to 2011. From 2011 to 2012 he held a Leverhulme Trust Early Career Fellowship and from 2012 to 2013 held a Newton International Fellowship, supported by the Royal Society and the British Academy, both at University College London (UCL). Since 2013, he is a Lecturer at UCL. He is a Core Team member of the European Space Agency (ESA) Planck satellite mission, a member of the Square Kilometre Array (SKA) Science Data Processor (SDP) design study, a member of the ESA Euclid satellite Science Consortium, and a member of the Large Synoptic Survey Telescope (LSST) Dark Energy Science Collaboration (DESC). His research interests include astroinformatics and astrostatistics, with a focus on signal and image processing, computational harmonic analysis, Bayesian analysis, and cosmology and radio interferometry.

\end{IEEEbiography}

\end{document}